\documentclass{amsart}
\usepackage{amssymb}
\usepackage{latexsym}

\date{\today}

\newcommand{\Z}{{\mathbb Z}}
\newcommand{\R}{{\mathbb R}}

\newcommand{\bbR}{{\mathbb{R}}}

\newcommand{\bbS}{{\mathbb{S}}}

\newtheorem{theorem}{Theorem}

\newtheorem{lemma}{Lemma}[section]
\newtheorem{prop}[lemma]{Proposition}
\newtheorem{coro}[lemma]{Corollary}

\DeclareMathOperator{\supp}{supp}

\newcounter{smalllist}
\newenvironment{SmallList}{%
\begin{list}{{\hbox to 1.5em{\hss\rm(\alph{smalllist})\hss}}}%
{\setlength{\topsep}{0mm}\setlength{\parsep}{0mm}\setlength{\itemsep}{0mm}%
\setlength{\labelwidth}{10mm}\usecounter{smalllist}}%
}{\end{list}}
\makeatletter\renewcommand{\thesmalllist}{\@alph\c@smalllist}\makeatother

\begin{document}

\title[Half-line Schr\smash{\"o}dinger Operators With No Bound States]{Half-line
Schr\"odinger Operators\\With No Bound States}
\author[D.\ Damanik and R.\ Killip]{David Damanik$^{1,2}$ and
Rowan Killip$^1$}
\thanks{$^1$ Department of Mathematics 253-37, California Institute of Technology,
Pasadena, CA 91125.
E-mail: damanik@its.caltech.edu; killip@its.caltech.edu}
\thanks{$^2$ Supported in part by NSF grants DMS-0227289 and INT-0204308}
\date{\today}

\begin{abstract}
We consider Sch\"odinger operators on the half-line, both discrete
and continuous, and show that the absence of bound states implies
the absence of embedded singular spectrum. More precisely, in the
discrete case we prove that if $\Delta + V$ has no spectrum
outside of the interval $[-2,2]$, then it has purely absolutely
continuous spectrum. In the continuum case we show that if both
$-\Delta + V$ and $-\Delta - V$ have no spectrum outside
$[0,\infty)$, then both operators are purely absolutely
continuous. These results extend to operators with finitely many
bound states.
\end{abstract}

\maketitle

\section{Introduction}

We study half-line Schr\"odinger operators, both continuous and
discrete, with a Dirichlet boundary condition at the origin.  That
is,
\begin{equation}\label{E:discop}
[h_V \psi] (n) = \psi(n+1) + \psi(n-1) + V(n)\psi(n)
\end{equation}
acting in $\ell^2(\Z^+)$, $\Z^+ = \{1,2,\ldots \}$, where
$\psi(0)=0$; and, in the continuum case,
\begin{equation}\label{E:contop}
[H_V \psi] (x) = -\psi''(x) + V(x) \psi(x)
\end{equation}
acting in $L^2([0,\infty))$ with the boundary condition
$\psi(0)=0$.  For convenience, we require that the potential, $V$,
be uniformly locally square integrable. We write
$\ell^\infty(L^2)$ for the Banach space of such functions.

The free operators, that is, when $V\equiv0$, can be diagonalized
by the Fourier transform.  This shows that they have spectra
$[-2,2]$ and $[0,\infty)$, respectively, and that in both cases,
the spectrum is purely absolutely continuous.

For a general discrete operator, the mere fact that the spectrum
is contained in $[-2,2]$ forces it to be purely absolutely
continuous. This is our first main result:

\begin{theorem}\label{T:biggy}
A discrete half-line Schr\"odinger operator $h_V$ with spectrum
contained in $[-2,2]$ has purely absolutely continuous spectrum.
\end{theorem}

In fact, the proof shows that $[-2,2]$ is the essential support of
the absolutely continuous spectrum. That is, for every $S
\subseteq [-2,2]$ of positive Lebesgue measure, the spectral
projection associated to $S$ is non-zero.

For any $V$ that is positive, the continuum operator $H_V$ has
spectrum contained in  $[0,\infty)$. Consequently, one may
conclude little from this requirement about the spectrum or its
type: the spectrum may have gaps, as periodic potentials
demonstrate, the spectral type may be pure point, such as occurs
in random models \cite{gmp,ks2}, or even purely singular
continuous, as certain sparse potentials show \cite{kls,p}. By
treating $V$ and $-V$ symmetrically, we obtain the continuum
analogue of Theorem~\ref{T:biggy}.

\begin{theorem}\label{T:Cbiggy}
Suppose $V \in \ell^\infty(L^2)$. If the spectra of both $H_V$ and
$H_{-V}$ are contained in $[0,\infty)$, then both operators have
purely absolutely continuous spectrum.  Moreover,
$\sigma(H_V)=\sigma(H_{-V}) = [0,\infty)$.
\end{theorem}

It follows from our proof that the essential support of the
absolutely continuous spectrum is equal to $[0,\infty)$.

The reason that sign-definite potentials do not offer
counterexamples to Theorem~\ref{T:biggy} is that, in the discrete
case, the spectrum of the free operator has two sides. Positive
potentials can produce spectrum above $+2$, and similarly,
negative potentials can produce spectrum below $-2$. In fact, the
operators $h_{-V}$ and $-h_V$ are unitarily equivalent. The
intertwining unitary operator is given by
\begin{equation}\label{E:U}
[U \psi](n) = (-1)^n \psi(n).
\end{equation}
Therefore, $\sigma(h_V) \subseteq [-2,2]$ is equivalent to
$\sigma(h_V) \subseteq [-2,\infty)$ and $\sigma(h_{-V}) \subseteq
[-2,\infty)$. In this way, we see that Theorem~\ref{T:Cbiggy} is
the natural analogue of Theorem~\ref{T:biggy}.

It has been shown, in \cite{ks}, that the free operator is the
only discrete whole-line Schr\"odinger operator with spectrum
contained in $[-2,2]$. A more transparent proof of this fact was
given in \cite{dhks}. This second proof is based on the
construction, for $V \not\equiv 0$, of certain trial functions
$\psi$ such that
$$
\langle \psi, (h_V - 2) \psi \rangle + \langle U \psi, (- h_V - 2)
U \psi \rangle > 0
$$
with $U$ defined as in \eqref{E:U}. (This inequality clearly
implies that $h_V$ must have spectrum outside $[-2,2]$.)

Similarly, on the whole space in two dimensions, only the free
operator has spectrum contained in $[-4,4]$. The corresponding
statement fails in three or more dimensions. (For proofs, see
\cite{dhks}.) The validity of this result in one or two dimensions
and its failure in three or more dimensions is intimately
connected to certain well-known facts about Schr\"odinger
operators in $\R^d$; see \cite{Cwi,Lieb,Roz,s2}. For example, if
$V\not\equiv0$ is a non-positive, smooth, compactly supported
potential on $\R^d$, then for $d=1,2$, $-\Delta + \lambda V$ has
bound states (isolated eigenvalues) for any $\lambda>0$, while for
$d\geq3$, $-\Delta + \lambda V$ has no bound states for small
$\lambda$.

For operators on the half-line, however, there are non-zero
potentials for which $\sigma(h_V) \subseteq [-2,2]$. The family of
potentials $V(n) = \lambda(-1)^n/n$ was studied in \cite{dhks}. It
was shown that for $|\lambda| \le 1$, $h_V$ has spectrum $[-2,2]$,
while for $|\lambda|>1$, it has infinitely many eigenvalues
outside $[-2,2]$.

On the other hand, absence of bound states is known to place
fairly stringent restrictions on the potential. For example, it
was shown in \cite[Corollary~9.3]{ks} that the potential must be
square summable. Moreover, by \cite{dk} (or \cite{ks}) this
implies that the (essential support of the) absolutely continuous
spectrum of the operator fills $[-2,2]$. In particular, it permits
one to conclude that if $\sigma(h_V) \subseteq [-2,2]$, then
actually $\sigma(h_V) = [-2,2]$. By the example given above,
absence of bound states does not imply $V \in \ell^1$; however,
Theorem~\ref{T:Vestimates} in Section~\ref{S:4} shows that $V$
must be weak-$\ell^1$ and so $\ell^p$ for every $p
> 1$.

Further restrictions were derived in \cite{dhks}. For example, by
Theorem~5.2 of that paper, any potential $V$ that does not produce
bound states must satisfy the pointwise bound $|V(n)| \le 2
n^{-1/2}$. It was also shown that there exists a sequence of
potentials $V_{m}$ such that $h_{V_{m}}$ has spectrum $[-2,2]$ for
every $m$ and $m^{1/2} |V_{m}(m)| \to 1$. Our
Proposition~\ref{P:roottwo} shows that $|V(n)| \le \sqrt{2/n}$ and
that for each $n$, there is a potential that realizes this bound.

None of the estimates for $V$ given above permits us to conclude
that the spectrum on $[-2,2]$ is purely absolutely continuous (as
is the case if $V\in \ell^1$, for example).  Indeed, following
Theorem~\ref{T:Dspec}, we exhibit, for any $\lambda > 1$, a
potential of the form $V(n) = \lambda (-1)^n n^{-1} + O(n^{-2})$
for which zero is an eigenvalue. This example is essentially a
discrete analogue of the classic Wigner--von~Neumann construction
\cite{WvN}.  As the potential $V(n)=(-1)^n n^{-1}$ has no bound
states, we see that the thresholds for the appearance of
eigenvalues inside and outside $[-2,2]$ are the same.  For this
reason, it is imperative that we obtain tight estimates at each
step.

A second important realization is that the correct quantity to
estimate is not the potential, $V$, but rather its ``conditional
integral'', $\sum_{m=n}^\infty V(m)$. For example, for every
$\varepsilon > 0$, there is a potential with $|V(n)| \le
\varepsilon/n$ and an embedded eigenvalue \cite{ek,r}. However,
Theorem~\ref{T:Dspec} below shows that this is not the case if the
conditional integral of $V$ obeys such an estimate.

In the continuum case, absence of bound states does not imply that
the potential goes to zero. Indeed, given any increasing function
$h : \R^+ \to \R^+$, there is a potential $V$ such that $V(x_k)
\ge h(x_k)$ for some sequence $x_k \to \infty$, and yet both $H_V$
and $H_{-V}$ have no bound states. This follows from Theorems~2.2
and~A.1 in \cite{dhs}. For example, if $h(x) = e^x$, one may
choose
$$
V(x) = \frac{d}{dx} \, \frac{\sin (e^{2x})}{4x}.
$$

As a compromise between generality and simplicity, we have chosen
to restrict our attention to potentials that are uniformly locally
square integrable.

The methods we employ to prove Theorems~\ref{T:biggy}
and~\ref{T:Cbiggy} will allow us to prove the following stronger
results:

\begin{theorem}\label{T:biggyer}
If a discrete half-line Schr\"odinger operator has only finitely
many eigenvalues outside $[-2,2]$, then it has purely absolutely
continuous spectrum on $[-2,2]$.
\end{theorem}

\begin{theorem}\label{T:Cbiggyer}
Suppose $V \in \ell^\infty(L^2)$. If both $H_V$ and $H_{-V}$ have
only finitely many eigenvalues below energy zero, then both
operators have purely absolutely continuous spectrum on the
interval $[0,\infty)$.
\end{theorem}

(Once again, the essential support of the absolutely continuous
spectrum fills out the interval indicated.)

A Jacobi matrix is an operator of the form
$$
[J\psi](n) = a_n \psi(n+1) + a_{n-1} \psi(n-1) + b_n \psi(n)
$$
acting in $\ell^2(\Z^+)$. The first step in our analysis is to use
the connection between such operators with spectrum contained in
$[-2,2]$ and the theory of polynomials orthogonal on the unit
circle, which seems to have first been made by Szeg\H{o}
(c.f.~\cite{szego}). This is discussed in Section~\ref{S:2}. In
particular, it is proved that a Jacobi matrix has spectrum
contained in $[-2,2]$ if and only if its parameters, $a_n$ and
$b_n$, can be represented in terms of a sequence of numbers
$\gamma_n \in (-1,1)$ as described by equations \eqref{E:BofG} and
\eqref{E:AofG}. The coefficients $\gamma_n$ occur in the continued
fraction expansion of a certain function associated to the Jacobi
matrix and, in the orthogonal polynomial context, are known as the
Verblunsky coefficients.

Sturm oscillation theory gives an alternate criterion for a Jacobi
matrix, $J$, to have $\sigma(J) \subseteq [-2,2]$ in terms of the
behaviour of the generalized eigenfunctions at energies $\pm 2$.
We discuss this in Section~\ref{S:3} and, in particular, we
determine the relation between these eigenfunctions and the
Verblunsky coefficients. This is used to motivate the definition
of the continuum analogue of the Verblunsky coefficients in
Section~\ref{S:6} and also to prove that $\pm 2$ are not
eigenvalues. As the Verblunsky coefficients with even and odd
indices play distinct roles in the discrete case, our continuum
analogue consists of two functions: $\Gamma_{{\rm e}}$ and
$\Gamma_{{\rm o}}$.

A related but different continuum analogue of the Verblunsky
coefficients was introduced by Kre\u \i n in his studies of the
continuum analogue of polynomials orthogonal on the unit circle
\cite{k}. Specifically, his function $A$ is given by our
$\Gamma_{{\rm e}} - \Gamma_{{\rm o}}$. He did not consider the
individual functions, nor any other combination of them.

Sections~\ref{S:4} and~\ref{S:6} are devoted to deriving estimates
for the Verblunsky coefficients; they treat the discrete and
continuum cases, respectively. It is also proved that there can be
no eigenvalues at the edges of the spectrum.

As noted earlier, it is the conditional integral of the potential
which proves to be the right object to study in order to prove the
theorems presented above. In Sections~\ref{S:5} and~\ref{S:7}, we
prove that certain estimates on this conditional integral imply
that the spectrum on $(-2,2)$ (resp., $(0,\infty)$) is purely
absolutely continuous. This is the content of
Theorems~\ref{T:Dspec} and~\ref{T:Cspec}. As the conditional
integral of the potential is given, to a good approximation, by
the even Verblunsky coefficients, the estimates derived in
Sections~\ref{S:4} and~\ref{S:6} provide the necessary input to
these theorems.

In order to prove Theorems~\ref{T:Dspec} and~\ref{T:Cspec}, we
study solutions of the corresponding eigenfunction equations using
Pr\"ufer variables. On the one hand, we show that they may only
grow or decay at a very restricted rate, and on the other, that
they actually remain bounded except on a set of energies of zero
Hausdorff dimension. By the Jitomirskaya-Last version \cite{jl} of
subordinacy theory \cite{gp}, the slow growth/decay of the
solutions implies that the spectral measure assigns zero weight to
sets of zero Hausdorff dimension. Moreover, the set of energies
where all solutions are bounded supports no singular spectrum and,
as just noted, the complement of this set has zero Hausdorff
dimension. These two statements preclude the existence of embedded
singular spectrum. A similar two-step procedure was used by
Remling \cite{r} to show that potentials which are $o(1/n)$ do not
have embedded singular spectrum.

As outlined above, Theorem~\ref{T:biggy} follows from
Theorem~\ref{T:Vestimates}, which provides estimates for the
conditional integral of the potential (see Section~\ref{S:4});
Theorem~\ref{T:pm2}, which precludes eigenvalues at $\pm 2$ (see
Section~\ref{S:4}); and Theorem~\ref{T:Dspec}, which is a general
criterion for the absence of singular spectrum embedded in
$(-2,2)$ (see Section~\ref{S:5}). Theorem~\ref{T:Cbiggy} follows
in a similar fashion from Theorem~\ref{T:contconseq} of
Section~\ref{S:6} and Theorem~\ref{T:Cspec} of Section~\ref{S:7}.

To obtain Theorems~\ref{T:biggyer} and~\ref{T:Cbiggyer}, which
permit finitely many bound states, we use a truncation argument to
show that the potential must obey estimates similar to those
derived in the no-bound-state case; see Corollaries~\ref{C:fmbs}
and~\ref{C:Cfmbs}. These corollaries provide the input to
Theorems~\ref{T:Dspec} and~\ref{T:Cspec} and also show that there
are no eigenvalues at the edges of the spectrum.

\medskip

\noindent\textit{Acknowledgments.} We thank Barry~Simon for his
encouragement and both the KTH and the Mittag-Leffler Institute
for their hospitality in the fall of 2002. Particular thanks go to
Ari~Laptev for his efforts in connection with the special
programme on Partial Differential Equations and Spectral Theory.

\section{Verblunsky Coefficients for Jacobi Matrices}\label{S:2}

In this section we work in the more general setting of Jacobi
matrices. Namely, we consider operators $J$ acting in
$\ell^2(\Z^+)$ by
\begin{equation}\label{jacobi}
[J\psi](n) = a_n \psi(n+1) + a_{n-1} \psi(n-1) + b_n \psi(n)
\end{equation}
where $\psi(0)$ is to be regarded as zero. (Recall that
$\Z^+=\{1,2,\ldots\}$.) The coefficients $a_n$ are positive and
the $b_n$ real.  Both sequences are assumed to be bounded and so
$J$ defines a bounded self-adjoint operator.

The question we wish to address is the following: for which
sequences of coefficients is the spectrum of $J$ contained in
$[-2,2]$?  (Note that in the case $a_n\equiv 1$, this is exactly
the question of which discrete Schr\"odinger operators have no
bound states.) While the criterion we prove in this section is by
no means easy to check, it is the basis for almost all the
analysis that follows.

\begin{theorem}\label{T:Jac&Verb}
A Jacobi matrix with coefficients $a_n$, $b_n$ has spectrum
$\sigma(J)\subseteq[-2,2]$ if and only if there is a sequence
$\gamma_n\in(-1,1)$, $n\in\{0,1,\ldots\}$, that obeys
\begin{align}
b_{n+1}   &= (1-\gamma_{2n-1})\gamma_{2n} - (1+\gamma_{2n-1})\gamma_{2n-2} \label{E:BofG}\\
a_{n+1}^2 &= (1-\gamma_{2n-1})(1-\gamma_{2n}^2)(1+\gamma_{2n+1}).          \label{E:AofG}
\end{align}
{\rm (}Here $\gamma_{-1}=-1$ and the value of $\gamma_{-2}$ is
irrelevant since it is multiplied by zero.{\rm )}
\end{theorem}

Most of this section is devoted to an exposition of the background
material and the introduction of notation; the ``Proof of Theorem
\ref{T:Jac&Verb}'' appears at the very end. Little that is said in
this section is new save perhaps the style of
presentation/derivation. In particular, Theorem~\ref{T:Jac&Verb}
appears in Geronimus \cite[\S31]{ger}, although not exactly in the
form stated above.

The Geronimus proof of Theorem~\ref{T:Jac&Verb} employs the
relation between orthogonal polynomials on the circle and on the
interval $[-2,2]$ derived by Szeg\H{o} \cite[Theorem~11.5]{szego}.
Our proof is more closely related to continued fractions. The
Schur algorithm provides a transformation on measures; it is
therefore natural to ask what transformation it induces on Jacobi
matrices. This is the content of Proposition~\ref{P:Schur}, which
seems to be new.

The proof of Theorem~\ref{T:Jac&Verb} presented below is short and
self-contained; we feel our discussion would be incomplete without
it.

It is not difficult to show that the vector
$\delta_1\in\ell^2(\Z^+)$ with entries $\delta_1(n)=\delta_{1,n}$
($\delta_{n,m}$ denotes the Kronecker delta function) is cyclic
for $J$.  That is, $\{ J^n \delta_1 : n=0,1,\ldots\}$ spans the
Hilbert space. Consequently, the spectrum of $J$ is equal to the
support of the spectral measure associated to $\delta_1$, which we
will denote by $d\mu$.  In fact, cyclicity implies that $J$ is
unitarily equivalent to $g(x)\mapsto xg(x)$ in $L^2(d\mu)$.  As
$\ell^2(\Z^+)$ is infinite-dimensional, so must be $L^2(d\mu)$,
which is equivalent to saying that $d\mu$ cannot be supported on a
finite set.  In fact, there is a one-to-one correspondence between
compactly supported probability measures $d\mu$ on $\R$ that are
not supported by a finite set and the set of Jacobi matrices with
uniformly bounded $a_n>0$, $b_n\in\R$. Given $\mu$, the sequences
$a_n$, $b_n$ are exactly the coefficients of the recurrence
relation obeyed by the polynomials orthonormal with respect to
$d\mu$.  (This is obvious once one realizes that the unitary
mapping $\ell^2(\Z^+)\to L^2(d\mu)$ described above maps
$\delta_n$ to the orthonormal polynomial of degree $n-1$.)

We also wish to discuss the $m$-function associated to $J$, that
is, the (1,1) entry of the Green function:
$m_0(z)=\langle\delta_1|(J-z)^{-1}\delta_1\rangle$. Naturally,
this can also be expressed in terms of the measure $d\mu$:
$$
m_0(z) = \int \frac{1}{t-z} \,d\mu(t).
$$
(The zero subscript is for consistency with what follows.)

If it happens that $\supp(d\mu)\subseteq[-2,2]$ (equivalently,
$\sigma(J)\subseteq[-2,2]$), then it is possible to define a
measure $d\rho$ on $\bbS^1=\{\zeta:|\zeta|=1\}$ which is symmetric
with respect to complex conjugation and obeys
$$
\int g(t) \,d\mu(t) = \int g(\zeta+\zeta^{-1}) d\rho(\zeta)
$$
for any measurable function $g$. These conditions uniquely determine $d\rho$.
In particular, note that $\rho(\bbS)=\mu([-2,2])=1$.

Associated to each measure on the circle is a Carath\'eodory function
$$
F_0(\xi) = \int \frac{\zeta+\xi}{\zeta-\xi} \,d\rho(\zeta) =
(\xi-\xi^{-1}) \, m_0(\xi+\xi^{-1}),
$$
defined and analytic for $\xi$ in the unit disk.  Notice that
$F(0)=\rho(\bbS^1)=1$ and that, because $\rho$ is symmetric,
$F_0:(-1,1)\to\bbR$.

To each such Caratheodory function $F_0$ is associated a Schur
function, an analytic mapping from the unit disk into itself, by
$$
 F_0(\xi) = \frac{1+\xi f_0(\xi)}{1-\xi f_0(\xi)} \quad \text{i.e.,} \quad
 f_0(\xi) = \frac{1}{\xi}\,\frac{F_0(\xi)-1}{F_0(\xi)+1}.
$$
The analyticity of $f_0$ follows from the fact that $F_0(0)=1$. As
$F_0:(-1,1)\to\bbR$, the same is true of $f_0$.  Note also that $f_0$ cannot
be a finite Blaschke product; if it were, then $d\rho$, and hence $d\mu$,
would be supported on a finite set, namely $\{\zeta : f_0(\zeta)=1\}$.

Recall that the Schur algorithm \cite{Schur} gives a one-to-one
correspondence between the set of Schur functions that are not
finite Blaschke products and the set of complex sequences
$\gamma:\{0,1,2,\ldots\} \to \{z:|z|<1\}$.   It proceeds as
follows:
$$
 f_{n+1}(\xi) = \frac{1}{\xi}\,\frac{f_n(\xi)-\gamma_n}{1-\overline{\gamma}_n f_n(\xi)}
 \qquad \gamma_n = f_n(0).
$$
The coefficients $\gamma_n$ have many names; following
\cite{simon}, we term them the Verblunsky coefficients.  (Other
common names are the Schur, Szeg\H o, Geronimus, or reflection
coefficients.)

As the measure, $d\rho$, we consider is symmetric with respect to
complex conjugation, so $f_0:(-1,1)\to(-1,1)$. It is easy to
verify inductively that this remains true for all $f_n$ and
consequently, that $\gamma_n\in(-1,1)$ for each
$n\in\{0,1,2,\ldots\}$.

\begin{lemma}\label{L:VJ}
The first two Verblunsky coefficients are
$$
\gamma_0 = \tfrac12 b_1 \quad\text{and}\quad \gamma_1 = - \frac{4
- b_1^2 - 2a_1^2}{4-b_1^2}.
$$
Equivalently, $b_1=2\gamma_0$ and $a_1^2 =
2(1-\gamma_0^2)(1+\gamma_1)$.
\end{lemma}

\begin{proof}
First,
$$
 \gamma_0 = f_0(0)=\tfrac12 F_0'(0) = \int \zeta^{-1} \,d\rho(\zeta) = \tfrac12 \int
 (\zeta+\zeta^{-1}) \,d\rho(\zeta)
$$
and so,
$$
 \gamma_0 = \tfrac12 \int t\,d\mu(t) = \tfrac12 b_1.
$$
For the second coefficient,
$$
 \gamma_1 = f_1(0) = \frac{f_0'(0)}{1-f_0(0)^2} = \frac{ F_0''(0) - \big[F_0'(0)\big]^2}{4-F_0'(0)}
$$
and
$$
  F''(0) = 4 \int \zeta^{-2} \,d\rho(\zeta) = \int 2t^2-4\,d\mu(t) = 2 \big( b_1^2+a_1^2\big) - 4
$$
which implies that
$$
  \gamma_1 = - \frac{4 - b_1^2 - 2a_1^2}{4-b_1^2}
$$
as claimed.
\end{proof}

The process by which $d\rho$ determines the Schur function $f_0$
may be inverted and so each of the iterates $f_n$ determines a
measure on $\bbS^1$.  The Carath\'eodory and $m$-functions of this
new measure will be denoted by $F_n$ and $m_n$, respectively; the
Jacobi matrix, by $J_n$. It turns out that there is a simple
relation between $J_2$, the Jacobi matrix resulting from two
iterations of the Schur algorithm, and the original matrix $J$.
Deriving this requires some computation. We begin by noting that
\begin{align*}
F_1(\xi)&=\frac{1+\xi f_1(\xi)}{1-\xi f_1(\xi)} = \frac{1-\gamma_0}{1+\gamma_0}\,\frac{1+f_0(\xi)}{1-f_0(\xi)}\\
&=
\frac{1-\gamma_0}{1+\gamma_0}\,\frac{(\xi+1)F_0(\xi)+(\xi-1)}{(\xi-1)F_0(\xi)+(\xi+1)}
\end{align*}
and that by iterating this,
\begin{align*}
F_2(\xi) =
\frac{1-\gamma_1}{1+\gamma_1}\,\frac{(\xi^2+1-2\gamma_0\xi)F_0(\xi)+(\xi^2-1)}
    {(\xi^2-1)F_0(\xi)+(\xi^2+1+2\gamma_0\xi)}.
\end{align*}
In this way, we obtain a relation between $m_2$ and $m_0$:
\begin{align}
m_2(z) &= \frac{1-\gamma_1}{1+\gamma_1}\,\frac{(z-2\gamma_0)m_0(z)
+ 1}{ (z^2-4)m_0(z) + (z+2\gamma_0)}
        \label{E:m2ofm0A} \\
&= \frac{4-b_1^2-a_1^2}{a_1^2} \;\frac{(z-b_1)m_0(z) + 1}{
(z^2-4)m_0(z) + (z+b_1)}, \label{E:m2ofm0B}
\end{align}
where we used the expressions for $\gamma_0$ and $\gamma_1$ given in the lemma above.

\begin{prop}\label{P:Schur}
If $\sigma(J)\subseteq[-2,2]$, then the Jacobi matrix resulting from two iterations of the Schur algorithm is
\begin{equation}\label{E:Jtwo}
J_2 = \begin{bmatrix}
 b &  a  &  0  &  0 \\
 a & b_3 & a_3 &  0 \\
 0 & a_3 & b_4 & \ddots  \\
 0 &  0  &\ddots& \ddots
\end{bmatrix}
\end{equation}
where $a$ and $b$ are determined by
\begin{align*}
\kappa^2 &= \tfrac{4-b_1^2}{4 - b_1^2 - a_1^2} = \tfrac{2}{1-\gamma_1} \\
a &= \kappa a_2 \\
b &= \kappa^2 b_2 + (\kappa^2 - 1) b_1 = \tfrac{2}{1-\gamma_1} b_2
+ 2\tfrac{1+\gamma_1}{1-\gamma_1}\gamma_0.
\end{align*}
Throughout, $a_n$, $b_n$ are the coefficients of the original Jacobi matrix.
\end{prop}

\begin{proof}
Let $J^{(j)}$ denote the matrix resulting from $J$ by the deletion
of the first $j$ rows and columns and let $m^{(j)}$ denote its
$m$-function.  For example,
$$
J^{(2)} =  \begin{bmatrix}
b_3 & a_3  &  0  &  0 \\
a_3 & b_4 & a_4 &  0 \vphantom{\ddots} \\
 0  & a_4 & b_5 & \ddots  \\
 0 &  0  &\ddots& \ddots
\end{bmatrix}.
$$

If $\tilde{m}$ denotes the $m$-function for the matrix $J_2$ of
(\ref{E:Jtwo}) then, by Cramer's rule,
\begin{equation}\label{E:tildem}
\tilde{m}(z) = \frac{1}{-z + b - a^2 m^{(2)}(z) }.
\end{equation}
Similarly,
$$
m^{(1)}(z) = \frac{1}{-z + b_2 - a_2^2 m^{(2)}(z) }
$$
and so $a_2^2 m^{(2)}(z)=(b_2-z) - [m^{(1)}(z)]^{-1}$.
Substituting this into (\ref{E:tildem}) gives
$$
\tilde{m}(z) = \frac{m^{(1)}(z)}{(1-\kappa^2)(-z - b_1) m^{(1)}(z)
+ \kappa^2}.
$$
We now use the fact that $a_1^2 m^{(1)}(z) = (b_1-z) -
[m_0(z)]^{-1}$ to obtain
\begin{align*}
\tilde{m}(z) &= \frac{(b_1-z)m_0 - 1}{(1-\kappa^2)(-z - b_1)[(b_1-z)m_0-1] + \kappa^2a_1^2m_0} \\
&= \frac{1}{\kappa^2-1} \; \frac{(z-b_1)m_0 + 1}{(z^2-4)m_0 +
(z+b_1)},
\end{align*}
where we also used $a_1^2\kappa^2=(4-b_1^2)(\kappa^2-1)$.  From
the definition of $\kappa$, this is exactly the same as the
expression for $m_2$ in terms of $m_0$ given in (\ref{E:m2ofm0B}).
Therefore $\tilde{m}=m_2$ and $m_2$ is the $m$-function for $J_2$.
Because a Jacobi matrix is uniquely determined by its
$m$-function, this proves \eqref{E:Jtwo}.
\end{proof}

\begin{coro}\label{C:gandab}
If the Jacobi matrix $J$ has $\sigma(J)\subseteq[-2,2]$, then the
corresponding Verblunsky coefficients $(\gamma_0,\gamma_1,\ldots)$
are related to the Jacobi matrix coefficients,
$(b_1,a_1,b_2,a_2,\ldots)$, by
\begin{align}
\gamma_{2n}  &= \tfrac{1}{1-\gamma_{2n-1}} b_{n+1} +
\tfrac{1+\gamma_{2n-1}}{1-\gamma_{2n-1}}\gamma_{2n-2}
    \label{E:gam2n} \\
\gamma_{2n+1}&=
(1-\gamma_{2n-1})^{-1}(1-\gamma_{2n}^2)^{-1}a_{n+1}^2 - 1
    \label{E:gam2n+1}
\end{align}
or, what is equivalent, by
\begin{align}
b_{n+1}   &= (1-\gamma_{2n-1})\gamma_{2n} - (1+\gamma_{2n-1})\gamma_{2n-2} \label{E:BofGa} \\
a_{n+1}^2 &= (1-\gamma_{2n-1})(1-\gamma_{2n}^2)(1+\gamma_{2n+1}). \label{E:AofGa}
\end{align}
In all formulae, $\gamma_{-1}=-1$.
\end{coro}

\begin{proof}
By iterating the proposition above, one finds that $m_{2n}$, the
$m$-function resulting from $2n$ iterations of the Schur
algorithm, is associated to the Jacobi matrix
$$
J_{2n} = \begin{bmatrix}
 b &  a  &  0  &  0 \\
 a & b_{n+2} & a_{n+2} &  0 \\
 0 & a_{n+2} & b_{n+3} & \ddots  \\
 0 &  0  &\ddots& \ddots
\end{bmatrix}
$$
where $a$ and $b$ are given by
$$
a^2=\frac{2a_{n+1}^2}{1-\gamma_{2n-1}} \qquad
b=\frac{2}{1-\gamma_{2n-1}}\Big\{
b_{n+1}+(1+\gamma_{2n-1})\gamma_{2n-2} \Big\}.
$$
Hence by Lemma~\ref{L:VJ},
$$
\gamma_{2n}  = \tfrac12 b \quad \text{and} \quad \gamma_{2n+1} =
\frac{2a^2}{4-b^2} - 1
$$
from which (\ref{E:gam2n}) and (\ref{E:gam2n+1}) follow by
substituting the formulae for $a$ and $b$ just given.
\end{proof}

\begin{proof}[Proof of Theorem \ref{T:Jac&Verb}]
If $\sigma(J)\subseteq[-2,2]$, then the corollary above shows that
the Verblunsky coefficients solve the equations \eqref{E:BofGa}
and \eqref{E:AofGa}, which are exactly the same as those stated in
the theorem. This proves one direction.

Given $J$, suppose there is a sequence
$(\gamma_0,\gamma_1,\ldots)$ with values in $(-1,1)$ so that both
\eqref{E:BofGa} and \eqref{E:AofGa} hold.  Then there is a Schur
function $f$ that has these coefficients and it must obey
$f(\bar\zeta)=\overline{f(\zeta)}$ because the coefficients are
real.  One may then define the corresponding $F$ and so a
probability measure $d\tilde\rho$ on $\bbS^1$ that is symmetric
with respect to complex conjugation.  This induces a probability
measure $d\tilde\mu$ on $[-2,2]$ which gives rise to a Jacobi
matrix, say $\tilde{J}$.  But, the coefficients of $\tilde{J}$ are
determined by the Verblunsky coefficients through
(\ref{E:BofGa})--\eqref{E:AofGa} and so must equal the
coefficients of $J$.  This implies $d\mu=d\tilde\mu$ and so $d\mu$
is supported in $[-2,2]$, which shows that
$\sigma(J)\subseteq[-2,2]$.
\end{proof}

\section{Verblunsky Coefficients and Eigenfunctions}\label{S:3}

Let $u$ and $w$ denote the generalized eigenfunctions at energies
$2$ and $-2$, respectively, with the standard normalization.  That
is,
\begin{equation}\label{uDef}
\begin{gathered}
  a_{n} u(n+1) + a_{n-1} u(n-1) + b_n u(n) = +2u(n) \\
  u(0)=0,\quad u(1)=1
\end{gathered}
\end{equation}
and
\begin{equation}\label{wDef}
\begin{gathered}
  a_{n} w(n+1) + a_{n-1} w(n-1) + b_n w(n) = -2w(n) \\
  w(0)=0,\quad w(1)=1.
\end{gathered}
\end{equation}
We also write $v(n)$ for $(-1)^{n-1}w(n)$, which obeys
\begin{equation}\label{vDef}
\begin{gathered}
  a_{n} v(n+1) + a_{n-1} v(n-1) - b_n v(n) = +2v(n) \\
  v(0)=0,\quad v(1)=1.
\end{gathered}
\end{equation}

Sturm oscillation theory for Jacobi matrices (see \cite{teschl})
shows that the Jacobi matrix, $J$, with coefficients $a_n$ and
$b_n$ (cf.~\eqref{jacobi}) has $\sigma(J)\subseteq[-2,2]$ if and
only if $u(n)$ and $v(n)$ are positive for all $n\in\Z^+$.  Hence
we have an alternative to the criterion discussed in the previous
section. However, we will not be using this alternate
characterization.

This section is devoted to discussing the relation between the
eigenfunctions $u,w$ and the Verblunsky coefficients, and so
provides a bridge between the two criteria. This serves two useful
purposes: (1) it simplifies the demonstration that $\pm2$ cannot
be eigenvalues of a discrete Schr\"odinger operator unless it has
spectrum outside $[-2,2]$; and (2) it motivates the definition of
the quantities that we regard as the continuum analogue of the
Verblunsky coefficients.

Not surprisingly, the values of $a_n$ and $b_n$ for
$1\leq n\leq N$ can be recovered from the values of $u(n)$ and $w
(n)$ for $2\leq n\leq N+1$.  The next lemma gives the precise formulae.

\begin{lemma}
If $W(n)=u(n+1)w(n)-u(n)w(n+1)$, the Wronskian of $u$ and $w$, and
$\tilde{W} (n) = u(n+1)w(n)+u(n)w(n+1)$, then
\begin{equation}\label{E:aofuw}
a_n = \frac{4}{W(n)} \sum_{k=1}^n u(k)w(k)
\end{equation}
\begin{equation}\label{E:bofuw}
b_n = \frac{-2}{u(n)w(n)}\left\{\frac{\tilde{W}(n)}{W(n)}
\sum_{k=1}^n u(k)w(k) + \frac{\tilde{W}(n-1)}{W(n-1)}
\sum_{k=1}^{n-1} u(k)w(k) \right\}.
\end{equation}
\end{lemma}

\begin{proof}
Consider multiplying (\ref{uDef}) by $w(n)$ and multiplying
(\ref{wDef}) by $u(n)$. Taking the difference gives
\begin{equation}\label{E:diff}
a_n W(n) - a_{n-1} W(n-1) = 4 u(n)w(n)
\end{equation}
while taking the sum gives
\begin{equation}\label{E:sum}
a_n \tilde{W}(n) + a_{n-1} \tilde{W}(n-1) + 2b_n  u(n)w(n) = 0.
\end{equation}
By summation, (\ref{E:diff}) implies
$$
a_n W(n) = 4 \sum_{k=1}^n u(k)w(k)
$$
which is equivalent to \eqref{E:aofuw}.  Having found the formula
for $a_n$ we can now solve \eqref{E:sum} for $b_n$.  This gives
\eqref{E:bofuw}.
\end{proof}

Similarly, one can write the Verblunsky coefficients in terms of
$u$ and $w$. The formulae actually look simpler than those for
$a_n$ and $b_n$:

\begin{lemma}
With $W$ and $\tilde{W}$ as in the previous lemma, we have
\begin{equation}\label{E:g2nofuw}
\gamma_{2n} = - \frac{\tilde{W}(n+1)}{W(n+1)}
\end{equation}
\begin{equation}\label{E:g2n+1ofuw}
\gamma_{2n+1} = -1 - \frac{2}{u(n+2)w(n+2)}\sum_{k=1}^{n+1}
u(k)w(k).
\end{equation}
\end{lemma}

\begin{proof}
We proceed by induction on $n$.  For $n=0$, we have
$$
\gamma_0 = \tfrac12 b_1 = - \frac{\tilde{W}(1)}{W(1)}
$$
by Corollary~\ref{C:gandab} and \eqref{E:bofuw}, respectively.
This implies that for $n=0$,
\begin{equation}\label{E:g2nblah}
\begin{split}
1-\gamma_{2n}^2 &= \frac{W(n+1)^2-\tilde{W}^2(n+1)}{W(n+1)^2} \\
&= -4 \frac{u(n+1)u(n+2)w(n+1)w(n+2)}{W(n+1)^2}.
\end{split}
\end{equation}
Using this together with Corollary~\ref{C:gandab} and
\eqref{E:bofuw} again,
$$
1+\gamma_1 = \tfrac12 a_1^2 (1-\gamma_0^2)^{-1} = -
\frac{2u(1)w(1)}{u(2)w(2)}.
$$

Now for the inductive step.  We assume that \eqref{E:g2nofuw},
\eqref{E:g2n+1ofuw}, and \eqref{E:g2nblah} hold and will show that
they remain true with $n$ replaced by $n+1$.

Using Corollary~\ref{C:gandab} and then \eqref{E:g2nofuw} and
\eqref{E:g2n+1ofuw},
\begin{align*}
\gamma_{2n+2} &= \frac{1}{1-\gamma_{2n+1}} \Big\{ b_{n+2} + (1 + \gamma_{2n+1})
\gamma_{2n} \Big\} \\
&= \frac{1}{1-\gamma_{2n+1}} \bigg\{ b_{n+2} +
\frac{2\tilde{W}(n+1)}{u(n+2)w(n+2)W(n+1)} \sum_{k=1}^{n+1}
u(k)w(k) \bigg\}.
\end{align*}
Continuing using (\ref{E:bofuw}) and \eqref{E:g2n+1ofuw} gives
\begin{align*}
\gamma_{2n+2} &= \frac{1}{1-\gamma_{2n+1}} \;
\frac{-2}{u(n+2)w(n+2)} \;
    \frac{\tilde{W}(n+2)}{W(n+2)} \sum_{k=1}^{n+2} u(k)w(k) \\
&= - \frac{\tilde{W}(n+2)}{W(n+2)}.
\end{align*}
Equation \eqref{E:g2nblah} with $n+1$ in lieu of $n$ follows easily from this.

After first employing \eqref{E:aofuw}, \eqref{E:g2n+1ofuw} together with what we have just
proved shows
\begin{align*}
1 + \gamma_{2n+3} &= (1-\gamma_{2n+1})^{-1}(1-\gamma_{2n+2}^2)^{-1}a_{n+2}^2 \\
&= \frac{-2}{u(n+3)w(n+3)} \sum_{k=1}^{n+2} u(k)w(k)
\end{align*}
just as is required to complete the proof.
\end{proof}

As we now demonstrate, the Verblunsky coefficients are intimately
related to the logarithmic derivatives of $u$ and $v$. In the case
where $a_n \equiv 1$, to which we will turn our attention shortly,
the odd and even coefficients are, to a good approximation, half
their sum and half their difference, respectively.

\begin{lemma}
Let $v(n) = (-1)^{n-1} w(n)$, as above, and write
\begin{equation}\label{discfandg}
F(n) = 1 - \frac{u(n+1)}{u(n+2)},\quad G(n) = 1 -
\frac{v(n+1)}{v(n+2)}
\end{equation}
for the logarithmic derivatives of $u$ and $v$. We have
\begin{equation}\label{gammasfromfg}
\gamma_{2n} = - \frac{F(n) - G(n)}{2 - F(n) - G(n)},\quad
\gamma_{2n+1} = - a_{n+1} \frac{F(n) + G(n)}{2} + a_{n+1} - 1
\end{equation}
and
\begin{equation}\label{fgfromgammas}
\begin{aligned}
F(n) &= a_{n+1}^{-1}\big\{ a_{n+1} - 1
    - \gamma_{2n+1} - \gamma_{2n} - \gamma_{2n+1}\gamma_{2n}\big\}, \\
G(n) &= a_{n+1}^{-1}\big\{ a_{n+1} - 1
    - \gamma_{2n+1} + \gamma_{2n} + \gamma_{2n+1}\gamma_{2n} \big\}.
\end{aligned}
\end{equation}
\end{lemma}

\begin{proof}
From \eqref{E:g2nofuw} we have
\begin{align*}
\gamma_{2n} &= - \frac{u(n+2) w(n+1) + w(n+2) u(n+1)}{u(n+2)
w(n+1) - w(n+2) u(n+1)}\\ &= - \frac{\left( 1 -
\frac{u(n+1)}{u(n+2)} \right) - \left( 1 - \frac{v(n+1)}{v(n+2)}
\right)}{2 - \left( 1 - \frac{u(n+1)}{u(n+2)} \right) - \left( 1 -
\frac{v(n+1)}{v(n+2)} \right)}.
\end{align*}
Combining \eqref{E:aofuw} and \eqref{E:g2n+1ofuw} gives
\begin{align*}
\gamma_{2n+1} &= -1 - \frac{a_{n+1}}{2} \left\{
\frac{w(n+1)}{w(n+2)} - \frac{u(n+1)}{u(n+2)} \right\}\\ &= -
\frac{a_{n+1}}{2} \left\{ \left( 1 - \frac{u(n+1)}{u(n+2)} \right)
+ \left( 1 - \frac{v(n+1)}{v(n+2)} \right) \right\} + a_{n+1} - 1.
\end{align*}
This proves \eqref{gammasfromfg}. The identities in
\eqref{fgfromgammas} are immediate consequences of
\eqref{gammasfromfg}.
\end{proof}

\section{Estimates for the Verblunksy Coefficients and the
Potential}\label{S:4}

Beginning with this section, we restrict ourselves to discrete
Schr\"odinger operators $h = \Delta + V$. That is, Jacobi matrices
with $a_n\equiv1$ and where we rename $b_n=V(n)$.

From Theorem \ref{T:Jac&Verb} we know that
$\sigma(h)\subseteq[-2,2]$ if and only if there exists a sequence
$\gamma_n$ with values in $(-1,1)$ that solves
\begin{align}
\label{b} V(n+1)  &= (1-\gamma_{2n-1})\gamma_{2n} - (1+\gamma_{2n-1})\gamma_{2n-2} \\
\label{a} 1 &= (1 - \gamma_{2n-1}) (1 - \gamma_{2n}^2) (1 + \gamma_{2n+1})
\end{align}
where $\gamma_{-1}=-1$ by definition. The proof shows that, when
it exists, the solution of this system is given by the Verblunsky
coefficients.

(We should also remind the reader that, as mentioned in the
introduction, for discrete Schr\"odinger operators,
$\sigma(h)\subseteq[-2,2]$ is equivalent to $\sigma(h)=[-2,2]$.)

The purpose of this section is to study the relations \eqref{b}
and \eqref{a}.  We will show that $\gamma_n$ must converge to zero
fairly rapidly and consequently so must $V(n)$.  We will also
show, by means of examples, that the decay estimates derived in
this section are optimal.

At the conclusion of the section, we show how estimates for the
potential derived under the assumption that there are no bound
states can be extended to the case where there are finitely many.

\begin{lemma}\label{oddupper}
If $h=\Delta + V$ is a discrete Schr\"odinger operator with
spectrum $[-2,2]$, then the associated Verblunsky coefficients
obey $\gamma_{2n-1}\le \gamma_{2n+1} \le 0$ for every $n \ge 1$.
That is, the odd Verblunsky coefficients are increasing and
non-positive.
\end{lemma}

\begin{proof}
Assume that, for some $n \ge 1$, we have $\gamma_{2n-1} > 0$.
Then, by \eqref{a},
$$
1 + \gamma_{2n+1} = (1 - \gamma^2_{2n})^{-1} (1 -
\gamma_{2n-1})^{-1} \ge (1 - \gamma_{2n-1})^{-1} \ge 1 +
\gamma_{2n-1}(1 + \gamma_{2n-1}).
$$
Iterating this, we obtain
$$
1 + \gamma_{2(n+m) + 1} \ge 1 + \gamma_{2n-1} (1 +
\gamma_{2n-1})^{m+1} \to \infty \; \mbox{ as } m \to \infty,
$$
which is a contradiction. Therefore, $\gamma_{2n-1} \le 0$ and
$$
1 + \gamma_{2n+1} \ge (1 - \gamma_{2n-1})^{-1} = 1 +
\frac{\gamma_{2n-1}}{1 - \gamma_{2n-1}},
$$
which yields $\gamma_{2n+1} \ge \gamma_{2n-1}$.
\end{proof}

\begin{lemma}\label{oddlower}
If $h = \Delta + V$ is a discrete Schr\"odinger operator with no
bound states and $\gamma_n$ are the associated Verblunsky
coefficients, then we have
\begin{equation}\label{E:oddlower}
\gamma_{2n+1} \ge - \frac{1}{n+2} + \sum_{j = 0}^n c_j^{(n)}
\gamma_{2j}^2
\end{equation}
for every $n \ge 0$, where
$$
c_j^{(n)} = \frac{(j+1)(j+2)}{(n+2)^2}.
$$
In particular, $\gamma_{2n+1} \ge \frac{-1}{n+2}$.
\end{lemma}

\begin{proof}
We proceed by induction on $n$. The case $n = 0$ follows from
$$
\gamma_1 = \tfrac{1}{2} (1 - \gamma_0^2)^{-1} - 1 \ge \tfrac{1}{2}
(1 + \gamma_0^2) - 1 = - \tfrac{1}{2} + \tfrac{1}{2} \gamma_0^2.
$$
For the induction step from $n-1$ to $n$, we note that
$$
1 - \gamma_{2n-1} \le 1 + \tfrac{1}{n+1} - \sum_{j = 0}^{n-1}
c_j^{(n-1)} \gamma_{2j}^2 = \tfrac{n+2}{n+1} - \sum_{j = 0}^{n-1}
c_j^{(n-1)} \gamma_{2j}^2,
$$
and hence
\begin{align*}
(1 - \gamma_{2n-1})^{-1} &\ge \Bigg( \tfrac{n+2}{n+1} -
    \sum_{j=0}^{n-1} c_j^{(n-1)} \gamma_{2j}^2 \Bigg)^{-1} \\
&= \tfrac{n+1}{n+2} \Bigg( 1 -
    \sum_{j = 0}^{n-1} \tfrac{n+1}{n+2}\, c_j^{(n-1)} \gamma_{2j}^2 \Bigg)^{-1} \\
&\ge \tfrac{n+1}{n+2} \Bigg( 1 + \sum_{j = 0}^{n-1}
    \tfrac{n+1}{n+2} \, c_j^{(n-1)} \gamma_{2j}^2 \Bigg).
\end{align*}
This yields
\begin{align*}
\gamma_{2n+1} & \ge (1 + \gamma_{2n}^2) \Bigg( \tfrac{n+1}{n+2} +
\sum_{j = 0}^{n-1} \tfrac{(n+1)^2}{(n+2)^2} \, c_j^{(n-1)}
\gamma_{2j}^2 \Bigg) - 1 \\
& \ge - \frac{1}{n+2} + \sum_{j = 0}^n c_j^{(n)} \gamma_{2j}^2
\end{align*}
since, for $0 \le j \le n-1$, $c_j^{(n)} = c_j^{(n-1)}
(n+1)^2/(n+2)^2$ and $c_n^{(n)} = (n+1)/(n+2)$.
\end{proof}

The first main result in this section is the determination of the
optimal pointwise estimate for potentials with no bound states.
Weaker results were obtained in \cite{dhks} by different methods.

\begin{prop}\label{P:roottwo}
If $\sigma(h_V)=[-2,2]$, then the potential obeys
\begin{equation}
    \big| V(n)\big| \leq \sqrt{\tfrac{2}{n}}.
\end{equation}
Moreover, for each $n$, there is a potential $V$ such that
$\sigma(h_V)\subseteq[-2,2]$ and $V(n)=\sqrt{2/n}$.
\end{prop}

\begin{proof}
The proof amounts to finding the sequence of $\gamma_j\in(-1,1)$ that maximizes
\begin{equation}\label{rtb}
V(n+1) = (1-\gamma_{2n-1})\gamma_{2n} - (1+\gamma_{2n-1})\gamma_{2n-2}
\end{equation}
subject to the constraint
\begin{equation}\label{rta}
    1 = (1 - \gamma_{2j-1}) (1 - \gamma_{2j}^2) (1 + \gamma_{2j+1}) \quad \text{for all $j$}.
\end{equation}
(The choice of $V(n+1)$ rather than $V(n)$ is to shorten the
subscripts in the equations that follow.) The existence of an
optimizer follows from the compactness of the set of
$(\gamma_0,\gamma_1,\ldots,\gamma_{2n+1})$ which obey \eqref{rta}
for $0\le j\le n$ and have $\gamma_{2n+1}\le0$. Note that the
final condition guarantees the possibility of extending this
sequence so that \eqref{rta} holds for all $j$.  For example, one
may choose $\gamma_j=0$ for $j\ge 2n+3$; the value of
$\gamma_{2n+2}$ being determined by \eqref{rta}.

As the sign of $\gamma_{2n}$ and of $\gamma_{2n-2}$ can be changed
without affecting the validity of \eqref{rta}, an optimizing
sequence must have $\gamma_{2n}\ge 0$ and $\gamma_{2n-2}\le 0$.

With $\gamma_{2n-2}$ and $\gamma_{2n-1}$ prescribed, $V(n+1)$ is
maximized by making $\gamma_{2n}$ as large as possible.  Choosing
$\gamma_j=0$ for $j\ge 2n+1$ and substituting this into
\eqref{rta} shows that
\begin{equation}\label{rt2n}
\gamma_{2n}^2 = 1-\frac{1}{1+\gamma_{2n-1}} = \frac{-\gamma_{2n-1}}{1-\gamma_{2n-1}}
\end{equation}
is possible.  In fact, since $\gamma_{2n+1}\le 0$ by
Lemma~\ref{oddupper}, this is also maximal.

Similarly, the optimizing $\gamma_{2n-2}$ obeys
\begin{equation}\label{rt2n-2}
\gamma_{2n-2}^2 = 1-\frac{1}{(1+\tfrac{1}{n})(1-\gamma_{2n-1})}
    = \frac{\tfrac{1}{n+1}+\gamma_{2n-1}}{1+\gamma_{2n-1}}.
\end{equation}
In this case, we wished to make $-\gamma_{2n-3}$ as large as
possible.  By Lemma~\ref{oddlower}, $\gamma_{2n-3}\geq
-\tfrac{1}{n}$.  This bound can be achieved by choosing
$\gamma_{2j}=0$ for $0\le j \le n-2$.

Combining the two preceding paragraphs, we see that we must find
the value of $\gamma_{2n-1}\in [-\frac{1}{n+1},0]$ that optimizes
$$
V(n+1) = \sqrt{  -\gamma_{2n-1}(1-\gamma_{2n-1})  \vphantom{\tfrac11}}
    + \sqrt{(\tfrac{1}{n+1}+\gamma_{2n-1})(1+\gamma_{2n-1})}.
$$
(This formula follows from substituting \eqref{rt2n} and
\eqref{rt2n-2} into \eqref{rtb}.) The resulting calculus exercise
has solution $\gamma_{2n-1}=-\tfrac{1}{2n+1}$ which gives
$V(n+1)=\sqrt{2/(n+1)}$.
\end{proof}

In \cite{dhks} it was shown that $|V(n)| \le 2 n^{-1/2}$ and
examples were given showing that the power $\frac12$ is optimal.
As was also noticed in \cite{dhks}, this pointwise estimate does
not tell the full story. The optimizing potential has only three
non-zero values:
$$
V_n(n-1)=\sqrt{\tfrac{n}{2(n+1)^2}},\quad
V_n(n)=\sqrt{\tfrac{2}{n}},\quad V_n(n+1)=\sqrt{\tfrac{1}{2n}}.
$$
Theorem~\ref{T:Vestimates} below shows that potentials without
bound states must decay much more quickly in an averaged sense.
First we give two propositions describing the decay of the
Verblunsky coefficients.

\begin{prop}\label{evenupper}
Suppose $h = \Delta + V$ is a discrete Schr\"odinger operator with
spectrum $[-2,2]$ and $\gamma_n$ are the associated Verblunsky
coefficients. For every $n \ge 0$,
\begin{equation}\label{lineven}
\sum_{j=0}^n (j+1)(j+2) \gamma_{2j}^2 \le n+2.
\end{equation}
This implies that for each $\varepsilon>0$, $\sum (j+1)^{1-\varepsilon} \gamma_{2j}^2 < \infty$.
It also implies that
\begin{equation}\label{E:weakl1}
 \#\{ j : |\gamma_{2j}| \ge \lambda \} \le \frac{9}{\lambda}
\end{equation}
and so $(\gamma_{2j})$ is weak-$\ell^1$.
\end{prop}

\begin{proof}
The bound \eqref{lineven} follows from \eqref{E:oddlower} because, by
Lemma~\ref{oddupper}, $\gamma_{2n+1}\leq 0$.

For the first implication, let $c_n =(n+2)^{-1-\varepsilon}$, which is summable, then
$$
\sum_{n=0}^\infty c_n \ge \sum_{n=0}^\infty c_n \sum_{j=0}^n \tfrac{(j+1)(j+2)}{n+2} \gamma_{2j}^2
= \sum_{j=0}^\infty (j+1)(j+2) \gamma_{2j}^2 \sum_{n = j}^\infty (n+2)^{-2-\varepsilon}.
$$
This proves the result because
$$
     (j+2)\sum_{n = j}^\infty (n+2)^{-2-\varepsilon}
\geq (j+2)\int_{j+2}^\infty x^{-2-\varepsilon}\,dx
\geq \tfrac{1}{1+\varepsilon}(j+2)^{-\varepsilon}.
$$

To prove \eqref{E:weakl1}, let $N_\lambda =\#\{ j : |\gamma_{2j}|
\ge \lambda \}$.  Note that the Verblunsky coefficients lie in
$(-1,1)$, so we need only consider $\lambda < 1$.  Clearly,
\begin{equation}\label{lambdabase}
\#\{ 1 \le j+1 < \lambda^{-1} : |\gamma_{2j}| \ge \lambda \} \le \lambda^{-1}.
\end{equation}
Moreover, for every $k \ge 0$, we infer from \eqref{lineven} that
$$
\#\{ 2^k \lambda^{-1} \le j+1 < 2^{k+1} \lambda^{-1} : |\gamma_{2j}| \ge \lambda \}
    \cdot 2^{2k} \lambda ^{-2} \cdot \lambda^2
\le 2 \cdot 2^{k+1} \lambda^{-1},
$$
that is,
\begin{equation}\label{lambdak}
\#\{ 2^k \lambda^{-1} \le j+1 < 2^{k+1} \lambda^{-1} : |\gamma_{2j}| \ge \lambda \}
\le 2^{2-k} \lambda^{-1}.
\end{equation}
Combining \eqref{lambdabase} and \eqref{lambdak}, we obtain
$$
N_\lambda \le \lambda^{-1} + \sum_{k=0}^\infty 2^{2-k} \lambda^{-1} = 9\,\lambda^{-1}
$$
which is exactly \eqref{E:weakl1}.
\end{proof}

\begin{prop}\label{evenupper2}
Suppose $h = \Delta + V$ is a discrete Schr\"odinger operator with
spectrum $[-2,2]$ and $\gamma_n$ are the associated Verblunsky
coefficients. There is a constant $C$ such that, for every $n\ge
1$,
\begin{equation}\label{logboundoneven}
\sum_{j=0}^n (j+1) \gamma_{2j}^2 \le \tfrac14 \log (n) + C
\end{equation}
and consequently,
\begin{equation}\label{logboundoneven2}
\sum_{j=0}^n |\gamma_{2j}| \le \tfrac12 \log (n) + C.
\end{equation}
\end{prop}

\begin{proof}
From \eqref{a} we have the identity
$$
\bigg(\frac{1 + \gamma_{2j-1}}{1 - \gamma_{2j-1}}\bigg)
\bigg(\frac{1 - \gamma_{2j+1}}{1 + \gamma_{2j+1}}\bigg)
= (1 - \gamma_{2j-1}^2) (1 - \gamma_{2j}^2)^{2} (1 - \gamma_{2j+1}^2).
$$
Applying the function $x\mapsto-\frac{1}{2}\log(x)$ to both sides and expanding in Taylor series yields
$$
(\gamma_{2j+1} - \gamma_{2j-1})\left\{ 1 + \sum_{l=1}^\infty \tfrac{1}{2l+1}
\sum_{k=0}^{2l} \gamma_{2j+1}^k \gamma_{2j-1}^{2l-k} \right\}
= \sum_{l = 1}^\infty \tfrac{1}{2l} \big\{ \gamma_{2j-1}^{2l} + 2\gamma_{2j}^{2l} + \gamma_{2j+1}^{2l} \big\}.
$$
We now estimate the left-hand side from above using the following consequence of Lemma~\ref{oddlower},
\begin{equation*}
1 + \sum_{l=1}^\infty \tfrac{1}{2l+1} \sum_{k=0}^{2l} \gamma_{2j+1}^k \gamma_{2j-1}^{2l-k}
\ge 1 + \sum_{l=1}^\infty (j+1)^{-2l} = \tfrac{(j+1)^2}{j (j+2)}
\end{equation*}
and estimate the right-hand side from below by neglecting all but the first term in the sum.
This gives
$$
\tfrac{(j+1)^2}{j (j+2)} (\gamma_{2j+1} - \gamma_{2j-1})
\geq \tfrac12\gamma_{2j-1}^{2} + \gamma_{2j}^{2} + \tfrac12\gamma_{2j+1}^{2}
$$
or, what is equivalent,
\begin{equation}\label{derivative}
\gamma_{2j+1} - \gamma_{2j-1} \ge \tfrac{j (j+2)}{2 (j+1)^2} \big(
\gamma_{2j-1}^2 + \gamma_{2j+1}^2 \big) + Y_j\quad\text{with}\quad Y_j=\tfrac{j (j+2)}{(j+1)^2}\gamma_{2j}^2.
\end{equation}

We now change variables according to
$$
\gamma_{2j-1} = \frac{\alpha_j}{j+1}
$$
so that $\alpha_0 = -1$ and, by Lemmas~\ref{oddupper} and \ref{oddlower}, $-1 \le \alpha_j \le 0$ for $j \ge 1$.
This implies
\begin{equation}\label{alphaj}
-\tfrac14 \le \alpha_j + \alpha_j^2 \le 0 \quad\text{for all $j$,}
\end{equation}
which we will use momentarily.  In the new variables, \eqref{derivative} reads
$$
\tfrac{(j + 3/2)}{(j+1)(j+2)} [ \alpha_{j+1} - \alpha_j ] -
\tfrac{1}{2(j+1)(j+2)} [ \alpha_{j+1} + \alpha_j ]
\leq
\tfrac{j(j+2)}{2(j+1)^2} \left[ \tfrac{\alpha_{j+1}^2}{(j+2)^2} +
\tfrac{\alpha_j^2}{(j+1)^2} \right] + Y_j.
$$
So, by rearranging terms and then using $\alpha_j^2\leq 1$ and $\alpha_{j+1}^2\leq 1$,
\begin{align}\label{alphader}
\alpha_{j+1} - \alpha_j
&\geq \tfrac{1}{2(j+3/2)} \left\{ \alpha_{j+1}
    + \tfrac{j}{j+1} \alpha_{j+1}^2 + \alpha_j
    + \tfrac{j(j+2)^2}{(j+1)^3} \alpha_j^2 \right\}
    + \tfrac{(j+1)(j+2)}{(j+ 3/2)} Y_j \\
&\geq \tfrac{1}{2(j+3/2)} \big\{ \alpha_{j+1} + \alpha_{j+1}^2 + \alpha_j + \alpha_j^2 \big\}
    - \tfrac{j+2}{(j+1)^3} + \tfrac{(j+1)(j+2)}{(j+ 3/2)} Y_j.
\end{align}
We now use \eqref{alphaj} and sum both sides to obtain
$$
\sum_{j = 0}^n \tfrac{(j+1)(j+2)}{(j+3/2)} Y_j \le \tfrac14 \log n + C,
$$
from which \eqref{logboundoneven} follows.

Equation \eqref{logboundoneven2} is an immediate consequence of
\eqref{logboundoneven} and the Cauchy--Schwarz inequality.
\end{proof}

We obtain the following corollaries for the potential $V(n)$.

\begin{theorem}\label{T:Vestimates}
If the discrete half-line Sch\"odinger operator $\Delta+V$ has
spectrum $[-2,2]$, then
\begin{SmallList}
\item the potential is weak-$\ell^1$ and so belongs to all $\ell^p$, $p>1$;

\item for all $\varepsilon>0$, $\sum n^{1-\varepsilon} |V(n)|^2 < \infty$;

\item there is a constant $C$ such that for all
$N\geq1$,\label{Tpartc}
$$
\sum_{n=1}^N |V(n)| \leq \log(N) + C;
$$

\item it is possible to write $V(n)=W(n)-W(n-1)+Q(n)$ with $Q\in \ell^1$, $W\in\ell^2$, and
$$
\sum_{n=1}^N n |W(n)|^2 \leq \tfrac{1}{4} \log(N) + C.
$$
\end{SmallList}
\end{theorem}

\begin{proof}  As $V(n+1)=(1-\gamma_{2n-1})\gamma_{2n} - (1+\gamma_{2n-1})\gamma_{2n-2}$,
parts (a)--(c) follow directly from the estimates on the
Verblunsky coefficients proved above.  For (d), we simply choose
$W(n)=\gamma_{2n-2}$, $n\geq1$, then
$Q(n)=-\gamma_{2n-3}(\gamma_{2n-2}+\gamma_{2n-4})$ for $n\geq 2$
and $Q(1)=\gamma_0$ which is summable.
\end{proof}

\noindent\textit{Example.} This example will show that all the
statements in Theorem~\ref{T:Vestimates} are optimal. This in turn
shows also that the estimates on the Verblunsky coefficients
obtained above (e.g., Propositions~\ref{evenupper} and
\ref{evenupper2}) are optimal.

Consider the potential $V(n) = (-1)^n/n$. It was shown in
\cite[Proposition~5.9]{dhks} that $\Delta + V$ has spectrum
$[-2,2]$. Consequently, weak-$\ell^1$ in (a) cannot be replaced by
$\ell^1$, (b) cannot be improved to $\sum n |V(n)|^2 < \infty$,
and the constant~$1$ in front of $\log(N)$ in (c) cannot be
decreased.

In order to see that this example also shows that the constant
$1/4$ appearing in (d) is the smallest possible, we note the
following:
$$
\log(N) - C \leq \sum_{n=1}^N |V(n)| \leq \sum_{n=1}^N |Q(n)| + 2
|W(n)|
$$
and so, since $Q\in\ell^1$, $\sum_{n=1}^N |W(n)| \geq
\frac{1}{2}\log(N) - C$.  By applying the Cauchy--Schwarz
inequality, we find $\sum_{n=1}^N n |W(n)|^2 \geq \tfrac14 \log(N)
- C$.

\medskip

Next we show that these estimates on the Verblunsky coefficients
allow for a short proof that $\pm2$ are not eigenvalues.

\begin{theorem}\label{T:pm2}
If $h$ is a discrete half-line Schr\"odinger operator with
spectrum $[-2,2]$, then $\pm2$ are not eigenvalues.
\end{theorem}

\begin{proof}
Of course, $E=2$ is an eigenvalue if and only if the generalized
eigenfunction at this energy, which we denote by $u$
(cf.~\eqref{uDef}), is square integrable. (We will concentrate on
$E=2$; $E=-2$ can be dealt with in the same manner.)

Since we are studying the Schr\"odinger operator case,
$a_n\equiv1$, equations \eqref{discfandg} and \eqref{fgfromgammas}
give us the following relation between $u$ and the Verblunsky
coefficients:
$$
\frac{u(n+1)}{u(n+2)} = 1 + \gamma_{2n+1} + \gamma_{2n} +
\gamma_{2n+1}\gamma_{2n}.
$$
Note that, by Sturm oscillation theory, $u(n)>0$, $n\geq1$, and
that, by definition, $u(1)=1$. Therefore, by neglecting the terms
$\gamma_{2n+1} \le 0$ and then using the summability of
$\gamma_{2j+1}\gamma_{2j}$,
\begin{align*}
\log\big[ u(n) \big] &= - \sum_{j = 0}^{n-2}
    \log\big[1 + \gamma_{2j+1} + \gamma_{2j} + \gamma_{2j+1}\gamma_{2j}\big]\\
& \ge C - \sum_{j = 0}^{n-2} \log\big[ 1 + \gamma_{2j} \big]
\end{align*}
for some constant $C$.  But, $\log[1+\gamma_{2j}]\leq |\gamma_{2j}|$ and so \eqref{logboundoneven2}
gives
$$
 \big|u(n)\big| \geq c n^{-1/2},
$$
which implies $u$ is not square summable.
\end{proof}

As intimated at the beginning of this section, the last two
theorems can be extended to the case of finitely many bound states
outside $[-2,2]$.

\begin{coro}\label{C:fmbs}
If the spectrum of a discrete half-line Schr\"odinger operator
$\Delta + V$ contains only finitely many points outside $[-2,2]$,
then
\begin{SmallList}
\item the potential is weak-$\ell^1$ and so belongs to all
$\ell^p$, $p>1$;

\item for all $\varepsilon > 0$, $\sum n^{1-\varepsilon} |V(n)|^2
< \infty$;

\item there is a constant $C$ such that for all
$N\geq1$,\label{Tpartcfb}
$$
\sum_{n=1}^N |V(n)| \leq \log(N) + C;
$$

\item it is possible to write $V(n)=W(n)-W(n-1)+Q(n)$ with $Q\in
\ell^1$, $W\in\ell^2$, and
$$
\sum_{n=1}^N n |W(n)|^2 \leq \tfrac{1}{4} \log(N) + C;
$$

\item $\pm2$ are not eigenvalues.

\end{SmallList}
\end{coro}

\begin{proof}
As $h_V$ has only finitely many eigenvalues outside $[-2,2]$, the
solutions $u$ and $v$, as defined in \eqref{uDef} and
\eqref{vDef}, pass through zero only finitely many times. (This
follows from the discrete analogue of the classical Sturm theory
\cite{teschl}.) So we may choose $k \in \Z^+$ such that $u(n)$ and
$v(n)$ do not change sign for $n \ge k$.

Using Sturm theory again, we see that the operator with potential
$V_1(n)=V(n+k)$ has no bound states. Thus, parts~(a)--(d) are
immediate consequences of Theorem~\ref{T:Vestimates}.

To prove (e), we will simply show that $u$ cannot be square
summable. Similar arguments show that the same is true of $v$.
This implies that $\pm2$ are not eigenvalues.

The sequence $\tilde{u}(n) = u(n+k)$ is the generalized
eigenfunction at energy $+2$ for the operator with potential
$$
V_2(n) = V(n+k) + \frac{u(k)}{u(k+1)}\delta_{n,1}.
$$

As $\Delta+V_1$ has no eigenvalues outside $[-2,2]$ and
$u(k)/u(k+1)$ is positive, $\Delta+V_2$ cannot have eigenvalues
below $-2$.  It also has no eigenvalues above $+2$; this is
because $\tilde u$, the generalized eigenfunction at energy $+2$,
does not pass through zero.

We have just seen that $\sigma(\Delta+V_2)\subseteq[-2,2]$;
therefore, by Theorem~\ref{T:pm2}, $\tilde u$ is not square
summable.  Consequently, $u$ is not square summable either.
\end{proof}

\section{Absence of Singular Spectrum: The Discrete
Case}\label{S:5}

The purpose of this section is to complete the proofs of
Theorems~\ref{T:biggy} and \ref{T:biggyer}. We have already seen,
in the previous section, that $\pm2$ are not eigenvalues, so it
suffices to consider $(-2,2)$.  Absence of singular spectrum in
this interval is a consequence of the following general result
whose applicability is guaranteed by part (d) of
Theorem~\ref{T:Vestimates} or by the same part of
Corollary~\ref{C:fmbs}.

\begin{theorem}\label{T:Dspec}
A half-line Schr\"odinger operator whose potential admits the decomposition
$V(n)=W(n)-W(n-1)+Q(n)$ with $Q\in\ell^1$, $W\in\ell^2$, and
\begin{equation}\label{E:West}
\sum_{n=1}^N n |W(n)|^2 \leq \tfrac{1}{4}\log(N) + C
\end{equation}
has purely absolutely continuous spectrum on the interval $(-2,2)$.
\end{theorem}

Of course, \eqref{E:West} implies $W(n)\leq C' \log(n)/n$ for
$n>1$ and so the assumption that $W\in\ell^2$ is redundant.

\medskip

\noindent\textit{Example.}  Define $\psi:\Z^+ \to \R$ as follows:
the absolute value is given by $|\psi(n)| = n^{-\alpha}$ and the
sign depends on the value of $n$ mod $4$ with the pattern
$+,+,-,-,\ldots$. If $\alpha>1/2$, then $\psi$ is square summable
and so a zero-energy eigenfunction for the potential
$$
V(n) = -\frac{\psi(n+1) + \psi(n-1)}{\psi(n)}
$$
for $n\geq2$ and $V(1)=-\psi(2)/\psi(1)$.  As $V(n) =
-2\alpha(-1)^n/n + O(n^{-2})$, the argument from the example
following Theorem~\ref{T:Vestimates} shows that any decomposition
$V(n) = W(n) - W(n-1) + Q(n)$ with $Q \in \ell^1$ has
$$
\sum_{n=1}^N n |W(n)|^2 \ge \alpha^2 \log(N) - C.
$$
Consequently, the constant $1/4$ in \eqref{E:West} cannot be
improved.

\medskip

The proof of Theorem~\ref{T:Dspec} will consume the remainder of
this section. As this requires a number of technical ingredients,
we first explain how the propositions that follow combine to
establish the result.

\begin{proof}[Overview of Proof]
The strategy we adopt to prove this theorem is inspired by
Remling's proof of absence of embedded singular spectrum for
$o(n^{-1})$ potentials \cite{r}. The method consists of two steps,
both combining the study of solutions to
\begin{equation}\label{E:psiE}
\psi(n+1) + \psi(n-1) + V(n) \psi(n) = E \psi(n)
\end{equation}
(with general initial conditions) with subordinacy theory.

First, we derive power-law estimates for all solutions of the
Schr\"odinger equation. These results are contained in
Proposition~\ref{P:powerlaw}.  It is shown that there are no
non-zero $\ell^2$ solutions for any $E\in(-2,2)$ and so no
embedded point spectrum.  Further, it is shown that, for
$E\in(-2,0)\cup(0,2)$, all non-zero solutions $\psi$ obey
$$
c n^{-3/5} \leq \bigl|\psi(n)\bigr|^2 + \bigl|\psi(n+1)\bigr|^2 \leq C n^{3/5}.
$$
By the Jitomirskaya--Last extension \cite{jl} of subordinacy
theory \cite{gp}, one may deduce that the restriction of the
spectral measure $d\mu$ to $(-2,2)$ gives zero weight to sets of
Hausdorff dimension less than $2/5$.  As noted a moment ago,
$d\mu$ gives zero weight to single points---this is why we could
write $(-2,2)$ in the last sentence rather than just
$(-2,0)\cup(0,2)$.

Second, we show that for all energies in $(-2,2)$ that lie outside
a set of zero Hausdorff dimension, all solutions of the
Schr\"odinger equation are bounded.  This is
Proposition~\ref{P:boundedsol}. By the most-used result of
subordinacy theory, this implies that any embedded singular
spectrum must be supported on this set of zero dimension and
$[-2,2]$ is contained in the essential support of the absolutely
continuous spectrum (see, e.g., \cite{s,stolz}).

Combining the preceding paragraphs, we see that on $(-2,2)$, the
singular part of the spectral measure must be supported by a set
of zero Hausdorff dimension, but also gives zero weight to sets of
zero dimension. Of course, only the zero measure gives no weight
to its support, so we may conclude that there is no embedded
singular spectrum.
\end{proof}

As just described, we need to study solutions of
\begin{equation}\label{E:Dsol}
\psi(n+1) + \psi(n-1) + V(n) \psi(n) = 2\cos(k)\,\psi(n)\qquad n\geq1
\end{equation}
where $\psi(0)$ is free to be anything---recall that the
generalized eigenfunction vanishes at $n=0$.  The parametrization
of energy $E\in(-2,2)$ as $2\cos(k)$, $k\in(0,\pi)$, is standard
and simplifies some of the formulae that follow.

Following \cite{kls,r}, we write $\psi(n)$ in terms of Pr\"ufer
variables $R$, $\theta$:
$$
\frac{1}{\sin k}
\begin{pmatrix} \phantom{-}\sin k & 0 \\ -\cos k & 1 \end{pmatrix}
\begin{pmatrix} \psi(n-1) \\ \psi(n) \end{pmatrix}
= R(n)
\begin{pmatrix} \sin\big(\theta(n)/2\big) - k \\ \cos\big(\theta(n)/2\big) - k \end{pmatrix}.
$$
These new variables obey the following equations:
\begin{align*}
\frac{R(n+1)^2}{R(n)^2} &= 1 - \frac{V(n)}{\sin k} \sin \theta(n)
    + \frac{V(n)^2}{\sin^2 k} \sin^2 \tfrac12 \theta(n), \\
\cot \big( \tfrac12\theta(n+1) - k\big) &= \cot\big( \tfrac12\theta(n)\big)
    - \frac{V(n)}{\sin k}.
\end{align*}
In the second equation, both sides being infinite is also permitted.
From here, Taylor expansion yields
\begin{align}
2 \log \left[ \frac{R(n+1)}{R(n)} \right] &= - V(n)\frac{\sin \theta(n)}{\sin k} + O( V(n)^2) \label{E:PR} \\
\theta(n+1) - \theta(n) &= 2k + \frac{V(n)}{\sin k} \left[ 1 -
\cos \theta (n) \right] + O(V(n)^2), \label{E:PT}
\end{align}
where the constants in the $O$-terms depend on $k$, but are independent of $n$.

As $V\in\ell^2$, \eqref{E:PR} gives the following two-sided bound on solutions of \eqref{E:Dsol}:
\begin{equation}\label{E:SolEstimate}
 \Bigl| \log \left[ |\psi(N+1)|^2 + |\psi(N)|^2 \right] \Bigr|
 \leq \frac{1}{\sin k} \biggl|\sum_{n=1}^N V(n)\sin \theta(n)\biggr| + C.
\end{equation}
(Note that by definition, $R(n)$ is comparable to the norm of the
vector $[\psi(n-1),\psi(n)]$.) This shows that in order to control
the behaviour of solutions, we must estimate $\sum
V(n)\sin\theta(n)$. Naturally, the first step is to invoke the
representation of $V$ in terms of $W$ and $Q$.  Using
\begin{align*}
\sin \theta(n) - \sin \theta(n+1) &= 2 \cos \left( \tfrac{\theta(n+1) + \theta(n)}{2} \right)
    \sin \left( \tfrac{\theta(n+1) - \theta(n)}{2} \right) \\
&= 2\cos \left( \tfrac{\theta(n+1) + \theta(n)}{2} \right)
    \sin \big( k + O(V(n)) \big) \\
&= 2\sin(k) \cos \left( \tfrac{\theta(n+1) + \theta(n)}{2} \right) + O\big(V(n)\big)
\end{align*}
together with $Q\in\ell^1$ and $W\in\ell^2$ yields
\begin{equation}\label{E:hohum}
\begin{aligned}
\sum_{n=1}^N V(n) \sin \theta(n)
&= \sum_{n=1}^N W(n) \big[ \sin\theta(n) - \sin\theta(n+1) \big] + O(1)\\
&= 2\sin k \sum_{n=1}^N W(n) \cos\left(\tfrac{\theta(n+1) + \theta(n)}{2}\right) + O(1),
\end{aligned}
\end{equation}
where, as before, the implicit constants depend on $k\in(0,\pi)$.

Combining \eqref{E:SolEstimate} and \eqref{E:hohum} shows that for each $E\in(-2,2)$,
\begin{equation}\label{E:SolAndW}
  \Bigl| \log \left[ |\psi(N+1)|^2 + |\psi(N)|^2 \right] \Bigr|\
  \le 2 \sum_{n=1}^N W(n) \cos\left(\tfrac{\theta(n+1) + \theta(n)}{2}\right)
  + C.
\end{equation}
Note how the gain of a factor $\sin(k)$ is important; it cancels
the $\frac{1}{\sin k}$ factor in front of the sum in
\eqref{E:SolEstimate}.  This is why estimates on $W$, the
``indefinite integral'' of $V$, control the behaviour of solutions
uniformly in energy. Estimates of the form
$$
\sum_{n=1}^N n |V(n)|^2 \le \alpha \log N + C
$$
do not preclude embedded eigenvalues, no matter how small one
chooses $\alpha$; see \cite{ek,r}.

\begin{lemma}\label{L:cos}
Given a sequence obeying $\phi(n+1)-\phi(n)=2k+o(1)$ for some
$k\in(0,\pi/2)\cup (\pi/2,\pi)$ and an $\varepsilon>0$, there is a
constant $C$ so that
\begin{equation}\label{E:cosL}
\sum_{n=1}^N \frac{\cos^2 \phi(n)}{n} \leq \bigl[\tfrac12+\varepsilon\bigr]\log(N) + C.
\end{equation}
\end{lemma}

\begin{proof}
By writing $\cos^2\phi=\frac12+\frac12\cos2\phi$ it suffices to show that
$\sum \cos[2\phi(n)]/n \leq  \varepsilon\log(N) + C$.

Recall the following estimate for the Dirichlet kernel
$$
\sup_\delta \Biggl\{ \sum_{j=0}^{\ell-1} \cos(4kj+\delta) \Biggr\} = \Biggl| \sum_{j=0}^{\ell-1}
e^{4ikj} \Biggr| = \left| \frac{\sin(2k\ell)}{\sin(2k)} \right| \leq \frac{1}{\left| \sin 2k
\right|}.
$$
It follows that for fixed $\ell\geq 4\left|\varepsilon\sin(2k)\right|^{-1}$ and $n$ sufficiently
large, depending on $k$, $\ell$, and $\varepsilon$,
\begin{align*}
\Biggl| \sum_{j=0}^{\ell-1} \cos\bigl[\phi(n+j)\bigr] \Biggr|
&\leq \Biggl| \sum_{j=0}^{\ell-1} \cos\bigl[2kj+\phi(n)\bigr] \Biggr|
    + \sum_{j=0}^{\ell-1} \bigl|\phi(n+j)-\phi(n)-2kj\bigr|
\leq \tfrac12 \varepsilon\ell.
\end{align*}

To finish the proof, note that for $n$ sufficiently large,
$$
\Biggl| \sum_{j=0}^{\ell-1} \frac{\cos 2\phi(n+j)}{n+j} \Biggr|
\leq \frac1n \Biggr|  \sum_{j=0}^{\ell-1} \cos[2\phi(n+j)] \Biggr|
    + \sum_{j=0}^{\ell-1} \frac{j}{n(n+j)}
\leq \frac{\varepsilon\ell}{n}
$$
so that \eqref{E:cosL} follows by summing over $\ell$-sized blocks
and absorbing the contribution from the initial segment, where $n$
is not sufficiently large, into the constant $C$.
\end{proof}

\begin{prop}\label{P:powerlaw}
Suppose $V(n)=W(n)-W(n-1)+Q(n)$ with $Q\in\ell^1$ and $W\in\ell^2$
obeying \eqref{E:West}. Then, for $E\in(-2,2)$, all solutions
$\psi$ of \eqref{E:psiE} that are not identically zero obey
$$
n^{-1} \lesssim \bigl|\psi(n)\bigr|^2 + \bigl|\psi(n+1)\bigr|^2
\lesssim n.
$$
Moreover, for non-zero energies,
$$
n^{-\eta} \lesssim \bigl|\psi(n)\bigr|^2 + \bigl|\psi(n+1)\bigr|^2
\lesssim n^\eta
$$
for any $\eta>1/\sqrt{2}$.
\end{prop}

\begin{proof}
By \eqref{E:SolAndW} it suffices to show that
$$
\Biggl| \sum_{n=1}^N W(n) \cos \phi(n) \Biggr| \leq \alpha(k) \log(N) + O(1)
$$
where $\phi(n)=\frac12[\theta(n+1) + \theta(n)]$,
$\alpha(\pi/2)=1/2$, and $\alpha(k)=\eta$ when $k\neq\pi/2$.
Applying Cauchy--Schwarz gives
$$
\Biggl| \sum_{n=1}^N W(n) \cos \phi(n) \Biggr|^2
\leq \sum_{n=1}^N n\bigl|W(n)\bigr|^2 \ \cdot\ \sum_{n=1}^N \frac{\cos^2 \phi(n)}{n}.
$$

By assumption, $\sum_1^N n |W(n)|^2 \leq \frac14 \log(N) + C$ and
so the $k=\pi/2$ case is an immediate consequence of $\cos^2
\phi(n)\leq 1$.

The $k\neq\pi/2$ case follows because, by \eqref{E:PT},
\begin{equation}\label{E:dphi}
\phi(n+1)-\phi(n) = \tfrac12\big[\theta(n+2)-\theta(n)\big] = 2k +
o(1)
\end{equation}
so we can apply Lemma~\ref{L:cos}.
\end{proof}

We now set about showing that the set of $E\in(-2,2)$ for which
not all solutions of \eqref{E:psiE} are bounded is of zero
Hausdorff dimension.  We begin with a lemma modelled on Theorem
3.3 of \cite{kls}.

\begin{lemma}\label{klstool}
Suppose $V(n)=W(n)-W(n-1)+Q(n)$ with $W$ and $Q$ as above and fix
$k\in (0,\pi)$.  If
$$
\widehat{W}(k;n) \equiv \lim_{M \to \infty} \sum_{m=n}^M W(m) e^{2ikm}
$$
exists and obeys
\begin{equation}\label{E:kls}
\sum_{n=1}^\infty | W(n+j) \widehat{W}(k;n)| <  \infty \qquad
\forall j\in\{1,0,-1\},
\end{equation}
then all solutions of \eqref{E:Dsol} are bounded.
\end{lemma}

\begin{proof}
By \eqref{E:SolAndW}, it suffices to show that
$$
\sum_{n=1}^N W(n)\exp\Big\{ \tfrac i2\big[\theta(n+1) + \theta(n)\big]\Big\}
$$
is bounded for those $k$ for which \eqref{E:kls} holds.  Writing
$\phi(n)=\frac12[\theta(n+1) + \theta(n)]$, we have
\begin{align*}
  \sum_{n=1}^N W(n)e^{i\phi(n)}
&= \sum_{n=1}^N \Big[\widehat{W}(k;n) - \widehat{W}(k;n+1)\Big] e^{i\phi(n)-2ikn} \\
&= \sum_{n=2}^N \widehat{W}(k;n) \Big[e^{i\phi(n)} - e^{i\phi(n-1)+2ik} \Big]e^{-2ikn}  + O(1).
\end{align*}
But by \eqref{E:PT}, $|\phi(n)-\phi(n-1)-2k|\leq 2[|W(n+1)|+2|W(n)|+|W(n-1)|]/\sin(k) + e_n$
where $e_n$ is summable. The result now follows easily from the fact that $|e^{ix}-e^{iy}|\leq |x-y|$.
\end{proof}

To control $\widehat{W}$ we use the following result from harmonic
analysis. For a proof, see \cite[\S XIII.11]{Zygmund} or \cite[\S
V.5]{Carleson}.

\begin{lemma}\label{L:S-Z}
For each $\varepsilon\in(0,1)$ and every measurable function
$m:[0,\pi]\to\Z$,
$$
\Biggl\{ \int\ \Biggl| \sum_{n=0}^{m(k)}  c_n e^{-2ink} \Biggr| \,
d\nu(k) \Biggr\}^2 \lesssim \mathcal{E}_\varepsilon (\nu)
\sum_{n=0}^\infty n^{1-\varepsilon} \big|c_n\big|^2
$$
where $\mathcal{E}_\varepsilon$ denotes the $\varepsilon$-energy
of $d\nu$: $\mathcal{E}_\varepsilon (\nu) = \int \int
|\sin(x-y)|^{-\varepsilon} \, d\nu(x) \, d\nu(y)$.
\end{lemma}

Combining these lemmas gives the following proposition, which
completes the proof of Theorem~\ref{T:Dspec} as described in the
overview given above.

\begin{prop}\label{P:boundedsol}
Suppose $V(n)=W(n)-W(n-1)+Q(n)$ with $Q\in\ell^1$ and $W\in\ell^2$ obeying \eqref{E:West}.
There is a set $S\subseteq(-2,2)$ of zero Hausdorff dimension so that for all
$E\in(-2,2)\setminus S$, all solutions $\psi$ of \eqref{E:psiE} are bounded.
\end{prop}

\begin{proof}
By applying the Cauchy--Schwarz inequality to dyadic blocks, for example, we see that
\eqref{E:West} implies $n^{-\varepsilon/4}W(n)\in\ell^1$ for all $\epsilon>0$.  Combining this with
Lemma~\ref{klstool} shows that we need only prove that for all $\varepsilon>0$, the set of $k$ for
which $n^{\varepsilon/4} \widehat{W}(k;n)$ is unbounded is of Hausdorff dimension no more than
$\varepsilon$.

Let $m(k)$ be a measurable integer-valued function on $(0,\pi)$.
Because of \eqref{E:West}, Lemma~\ref{L:S-Z} implies
\begin{align*}
   \int\ \Biggl| \sum_{n=m_l(k)}^{2^{l+1} - 1} W(n) e^{2ikn} \Biggr| \, d \nu(k)
&= \int\ \Biggl| \sum_{n=0}^{\tilde{m}_l(k)} W(2^{l+1}-1-n) e^{-2ikn} \Biggr| \, d \nu(k) \\
&\lesssim \Biggl\{ \sum_{n=2^l}^{2^{l+1}-1} n^{1-\varepsilon}
\big|W(n)\big|^2 \Biggr\}^{1/2}
    \sqrt{\mathcal{E}_\varepsilon (\nu)} \\
&\lesssim \sqrt{l} \, 2^{-\varepsilon l/2}
\sqrt{\mathcal{E}_\varepsilon (\nu)}
\end{align*}
where $m_l(k)=\max\{m(k),2^l\}$, $\tilde{m}_l(k)=\min\{2^l-1,2^{l+1}-1-m(k)\}$, and sums with
lower index greater than their upper index are to be treated as zero.
Multiplying both sides by $2^{\varepsilon l/4}$, summing this over $l$,
and applying the triangle inequality on the left gives
$$
\int\ \Biggl| m(k)^{\varepsilon/4} \sum_{n=m(k)}^{\infty} W(n)
e^{2ikn} \Biggr| \, d \nu(k) \lesssim
\sqrt{\mathcal{E}_\varepsilon (\nu)}.
$$
That is, for any measurable integer-valued function $m(k)$,
$$
\int m(k)^{\varepsilon/4} \bigl| \widehat{W}(k;m(k))\bigr| \, d\nu
\lesssim \sqrt{\mathcal{E}_\varepsilon (\nu)}.
$$
This implies that the set on which $n^{\varepsilon/4} \widehat{W}(k;n)$ is unbounded must be of zero
$\varepsilon$-capacity  (i.e., it does not support a measure of finite $\varepsilon$-energy).

As the Hausdorff dimension of sets of zero $\varepsilon$-capacity is less than or equal to $\varepsilon$
(see \cite[\S IV.1]{Carleson}), this completes the proof.
\end{proof}

\section{A Continuum Analogue of the Verblunsky
Coefficients}\label{S:6}

As in the introduction, we write $H_V$ for the Schr\"odinger
operator associated to the potential $V$:
$$
[H_V \psi](x) = - \psi''(x) + V(x) \psi(x).
$$
We require a Dirichlet boundary condition at zero, $\psi(0)= 0$,
and $V \in \ell^\infty(L^2)$, that is,
\begin{equation}\label{potass}
\sup_{n \ge 0} \int_n^{n+1} |V(t)|^2 \, dt < \infty.
\end{equation}

The purpose of this section is to identify the continuum analogue
of the Verblunsky coefficients and to derive estimates for them.

It is well known that \eqref{potass} ensures that for every energy
$E$ and every boundary condition $\alpha$ at zero, there exists a
locally $H^2$ solution to
\begin{equation}\label{conteve}
-\psi''(x) + V(x) \psi(x) = E \psi(x), \; \psi(0) = \sin \alpha,
\; \psi'(0) = \cos \alpha.
\end{equation}
See, for example, \cite{w}.

Let $u$ and $v$ denote the zero-energy normalized Dirichlet
solutions of \eqref{conteve} with potential $V$ and $-V$,
respectively. That is,
\begin{equation}\label{E:uwDefn}
\begin{aligned}
 -u'' + V u &= 0 \quad u(0)=0,\ u'(0)=1 \\
 -v'' - V v &= 0 \quad v(0)=0,\ v'(0)=1.
\end{aligned}
\end{equation}
Notice that $u$ and $v$ play the same roles as they did in the
discrete case; compare \eqref{uDef} and \eqref{vDef}.

If both $H_V$ and $H_{-V}$ have no bound states, then it follows
from oscillation theory (see, e.g., \cite{cl,w}) that the
functions $u$ and $v$ have no zeros in $(0,\infty)$.

From $u$ and $v$ we define the two functions, $\Gamma_{{\rm e}}$
and $\Gamma_{{\rm o}}$, on $(0,\infty)$ in a manner inspired by
\eqref{discfandg} and \eqref{gammasfromfg}:
\begin{align*}
\Gamma_{{\rm e}}(x) &= \frac12 \left[ \frac{u'(x)}{u(x)} -
\frac{v'(x)}{v(x)} \right]\\
\Gamma_{{\rm o}}(x) &= - \frac12 \left[ \frac{u'(x)}{u(x)} +
\frac{v'(x)}{v(x)} \right].
\end{align*}
These two functions are the analogues of the Verblunsky
coefficients in the discrete case with even and odd index,
respectively. All the crucial properties of the $\gamma_{2n}$'s
and the $\gamma_{2n+1}$'s carry over to the continuum case, as we
will see. Lemma~\ref{L:6.1} below shows that they obey a pair of
differential equations which are the analogues of the formulae
\eqref{b} and \eqref{a}.

That the Verblunsky coefficients are related to the logarithmic
derivative of eigenfunctions in the discrete case appears in
Geronimus' work on orthogonal polynomials \cite[\S31]{ger}. In
Kre\u \i n's studies of a continuum analogue of polynomials
orthogonal on the unit circle (see, e.g., \cite{k}), he introduced
a function $A$ which plays the role of the Verblunsky
coefficients. In the case where $A$ is a real-valued function, it
is given by the logarithmic derivative of the $u$ associated with
the potential $A' + A^2$. In this way, $A = \Gamma_{{\rm e}} -
\Gamma_{{\rm o}}$. While the two approaches are related, the Kre\u
\i n approach is not suited to our problem. For an example of how
the Kre\u \i n approach may be employed in the study of
Schr\"odinger operators, see \cite{den}.

\begin{lemma}\label{L:6.1}
The functions $\Gamma_{{\rm e}},\Gamma_{{\rm o}}$ obey
\begin{align}
\label{gammaede} \Gamma_{{\rm e}}'(x) &= V(x) + 2 \Gamma_{{\rm e}}
(x) \Gamma_{{\rm
o}} (x) \\
\label{gammaode} \Gamma_{{\rm o}}'(x) &= \Gamma_{{\rm o}}^2 (x) +
\Gamma_{{\rm e}}^2 (x).
\end{align}
\end{lemma}

\begin{proof}
Write
\begin{equation}\label{fandg}
F(x) = \frac{u'(x)}{u(x)} \mbox{ and } G(x) = \frac{v'(x)}{v(x)}
\end{equation}
so that $\Gamma_{{\rm e}} (x) = [F(x) - G(x)]/2$ and $\Gamma_{{\rm
o}} (x) = - [ F(x) + G(x) ] /2$. We infer from differential
equations for $u$ and $w$, \eqref{E:uwDefn}, that
\begin{equation}\label{fgde}
F'(x) = V(x) - F^2(x) \mbox{ and } G'(x) = -V(x) - G^2(x).
\end{equation}
Subtraction gives
$$
F'(x) - G'(x) = 2V(x) - F^2(x) + G^2(x)
$$
and from this we get
$$
V(x) = \tfrac12 (F'(x)-G'(x)) + \tfrac12[(F(x) - G(x)][F(x) +
G(x)],
$$
which is \eqref{gammaede}. On the other hand, addition of the
identities in \eqref{fgde} yields
$$
-\tfrac12 (F'(x) + G'(x)) = \tfrac12 F^2(x) + \tfrac12 G^2(x) =
\left( \frac{F(x) - G(x)}{2} \right)^2 + \left( - \frac{F(x) +
G(x)}{2} \right)^2\!\!,
$$
which is \eqref{gammaode}.
\end{proof}

We will now present three lemmas, which are the continuum
analogues of results proved in Section~\ref{S:4}.  We begin with
the counterpart to Lemmas~\ref{oddupper} and~\ref{oddlower}.

\begin{lemma}\label{gammaoneg}
For every $x > 0$, we have
$$
- \frac1x \le \Gamma_{{\rm o}} (x) \le 0.
$$
\end{lemma}

\begin{proof}
Given $x_0 > 0$ and $y_0 \not=0$, consider the initial value
problem $\Gamma'(x) = \Gamma^2(x), \; \Gamma(x_0) = y_0$. Its
solution is given by
$$
\Gamma(x) = - \left( x - \frac{1 + x_0 y_0}{y_0} \right)^{-1}.
$$
Notice that if $y_0>0$, then $\Gamma$ blows up at finite $x>x_0$.

By \eqref{gammaode}, $\Gamma_{{\rm o}}'(x) \geq \Gamma_{{\rm
o}}^2(x)$.  Therefore,
\begin{equation}\label{whatever}
\Gamma_{{\rm o}}(x) \ge - \left( x - \frac{1 + x_0 \Gamma_{{\rm
o}}(x_0)}{\Gamma_{{\rm o}}(x_0)} \right)^{-1}
\end{equation}
for $x > x_0$. As $\Gamma_{{\rm o}}(x)$ is regular, blow-up cannot
occur and, by the remark made earlier, this implies that
$\Gamma_{{\rm o}}(x)\leq0$ for all $x\in(0,\infty)$.

By \eqref{whatever}, $\Gamma_{{\rm o}} (x) \ge - 1/x$ follows from
$$
\lim_{x_0 \to 0} \frac{1 + x_0 \Gamma_{{\rm o}}(x_0)}{\Gamma_{{\rm
o}}(x_0)} = 0,
$$
which in turn follows from
$$
\frac{1 + x_0 \Gamma_{{\rm o}}(x_0)}{\Gamma_{{\rm o}}(x_0)} = x_0
- 2 \left( \frac{u'(x_0)}{u(x_0)} + \frac{v'(x_0)}{v(x_0)}
\right)^{-1}
$$
and the fact that $u$, $u'$, $v$, and $v'$ are continuous at the
origin with the values given in \eqref{E:uwDefn}.
\end{proof}

In place of Proposition~\ref{evenupper} we have:

\begin{lemma}\label{gammaet2bound}
For every $x \ge 0$,
\begin{equation}\label{E:gammaet2}
\int_0^x t^2 \cdot \left[ \Gamma_{{\rm e}}^2 (t) + \left(
\Gamma_{{\rm o}} (t) + \tfrac1t \right)^2 \right] \, dt \le x.
\end{equation}
Moreover, $| \{ x  : |\Gamma_{{\rm e}} (x) | \ge \lambda \} | \le
 5 \lambda^{-1}$ and so $\Gamma_{{\rm e}}\in L^1_w$.
\end{lemma}

\begin{proof}
Write
$$
\Gamma_{{\rm o}}(t) = -\frac1t + h(t).
$$
It follows from the definition of $\Gamma_{{\rm o}}$ and
Lemma~\ref{gammaoneg} that
\begin{equation}\label{hprop}
0 \le h(t) \le \frac1t \mbox{ and } \lim_{t \to 0+} t^2 h(t) = 0.
\end{equation}
Differentiating the definition of $h$ gives
$$
\Gamma_{{\rm o}}'(t) = \frac{1}{t^2} + h'(t),
$$
while from \eqref{gammaode} we have
$$
\Gamma_{{\rm o}}'(t) = \Gamma_{{\rm o}}^2 (t) + \Gamma_{{\rm e}}^2
(t) = \frac{1}{t^2} - \frac{2 h(t)}{t} + h^2(t) + \Gamma_{{\rm
e}}^2(t).
$$
Therefore,
$$
h'(t) + \frac2t h(t) = h^2(t) + \Gamma_{{\rm e}}^2 (t),
$$
which in turn implies
$$
(t^2 h(t))' = t^2 h^2(t) + t^2 \Gamma_{{\rm e}}^2(t).
$$
The first inequality, \eqref{E:gammaet2}, now follows by
integrating this and applying \eqref{hprop}.

To prove the second estimate, notice that for $k \ge 0$,
\eqref{E:gammaet2} implies
$$
| \{ 2^k \lambda^{-1} \le x < 2^{k+1} \lambda^{-1} : |\Gamma_{{\rm
e}} (x) | \ge \lambda \} | \cdot 2^{2k} \le 2^{k+1} \lambda^{-1},
$$
which yields
$$
| \{ x > 0 : |\Gamma_{{\rm e}} (x) | \ge \lambda \} | \le
\lambda^{-1} + \sum_{k = 0}^\infty 2^{1-k} \lambda^{-1} = 5
\lambda^{-1},
$$
concluding the proof.
\end{proof}

Lastly, the continuum analogues of Proposition~\ref{evenupper2}
and part (\ref{Tpartc}) of Theorem~\ref{T:Vestimates} are given by
the following:

\begin{lemma}\label{gammaet1bound}
The function $\Gamma_{{\rm e}}$ admits the following estimates:
for all $x > y > 0$,
$$
\int_y^x t \cdot \Gamma_{{\rm e}}^2 (t) \, dt \le 1 + \tfrac14
\log \left(\tfrac{x}{y} \right)
$$
and for $x>1$,
\begin{equation}\label{ell1}
\int_1^x |\Gamma_{{\rm e}} (t)| \, dt \le \tfrac12 \log(x) + C.
\end{equation}
\end{lemma}

\begin{proof}
Write
$$
\Gamma_{{\rm o}} (t) = - \frac{\alpha(t)}{t}.
$$
By Lemma~\ref{gammaoneg}, we have $0 \le \alpha(t) \le 1$ for
every $t > 0$. From
$$
\Gamma_{{\rm o}}' (t) = - \frac{\alpha'(t)}{t} +
\frac{\alpha(t)}{t^2}
$$
and \eqref{gammaode} we obtain
$$
\Gamma_{{\rm e}}^2 (t) = \Gamma_{{\rm o}}' (t) - \Gamma_{{\rm
o}}^2 (t) = - \frac{\alpha'(t)}{t} + \frac{\alpha(t) -
\alpha^2(t)}{t^2}.
$$
Thus, because $0 \le \alpha(t) \le 1$ implies $0 \le \alpha(t) -
\alpha^2(t) \le \frac14$,
$$
\int_y^x t \cdot \Gamma_{{\rm e}}^2 (t) \, dt = \int_y^x  -
\alpha'(t) + \frac{\alpha(t) - \alpha^2(t)}{t} \, dt \le 1 +
\int_y^x \frac{1}{4t} \, dt,
$$
from which the first estimate follows.

The second estimate follows from the first by applying the
Cauchy--Schwarz inequality.
\end{proof}

The following theorem gives the input necessary to prove the
absence of singular spectrum in the next section. Specifically, it
shows that absence of bound states (for both $H_V$ and $H_{-V}$)
forces the potential to have a certain structure and so to be
amenable to treatment by the general criterion given in
Theorem~\ref{T:Cspec} below.

\begin{theorem}\label{T:contconseq}
If $V \in \ell^\infty(L^2)$ and the spectra of both $H_V$ and
$H_{-V}$ are contained in $[0,\infty)$, then
\begin{SmallList}
\item we can write $V = W' + Q$ with $Q \in L^1$, $W' \in
\ell^\infty (L^2)$, and
\begin{equation}\label{E:Wbound}
\int_1^x t [W(t)]^2 \le \tfrac14 \log(x) + 1;
\end{equation}
\item neither $H_V$ nor $H_{-V}$ has zero as an eigenvalue.
\end{SmallList}
\end{theorem}

\begin{proof}
(a) Let $g$ be a $C^\infty$ function on $\R^+$ with $g(x) = 0$ for
$0 \le x \le 1/2$ and $g(x) = 1$ for $x \ge 1$. Let $W(x) = g(x)
\Gamma_{{\rm e}} (x)$ and $Q = V - W'$. Thus, for $x \ge 1$, we
have $W(x) = \Gamma_{{\rm e}} (x)$ and, by \eqref{gammaede}, $Q(x)
= - 2 \Gamma_{{\rm e}} (x) \Gamma_{{\rm o}} (x)$. By
\eqref{potass}, $Q$ is absolutely integrable on $(0,1)$, and by
Lemmas~\ref{gammaoneg} and~\ref{gammaet2bound}, it is absolutely
integrable on $(1,\infty)$. Moreover, $W' \in \ell^\infty(L^2)$
follows from \eqref{potass}, Lemma~\ref{gammaoneg}, and
Lemma~\ref{gammaet2bound}. Finally, the bound \eqref{E:Wbound}
follows from Lemma~\ref{gammaet1bound}.

(b) From the definitions of $\Gamma_{{\rm e}}$ and $\Gamma_{{\rm
o}}$,
$$
\frac{u'(x)}{u(x)} = \Gamma_{{\rm e}} (x) - \Gamma_{{\rm o}} (x)
$$
and hence, by Lemma~\ref{gammaoneg},
$$
\log \bigl[u(x)\bigr] \ge C - \int_1^x |\Gamma_{{\rm e}}(t)| \, dt
$$
for $x>1$. By using \eqref{ell1}, we obtain
$$
u(x) \gtrsim x^{-1/2}
$$
for $x>1$. Therefore, $u\not\in L^2$ and so zero is not an
eigenvalue of $H_V$.  Similar reasoning shows that $H_{-V}$ does
not have zero as an eigenvalue.
\end{proof}

\begin{coro}\label{C:Cfmbs}
If $V \in \ell^\infty(L^2)$ and both $H_V$ and $H_{-V}$ have only
finitely many eigenvalues below zero, then
\begin{SmallList}
\item we can write $V = W' + Q$ with $Q \in L^1$, $W' \in
\ell^\infty (L^2)$, and
\begin{equation*}
\int_1^x t [W(t)]^2 \le \tfrac14 \log(x) + 1;
\end{equation*}
\item neither $H_V$ nor $H_{-V}$ has zero as an eigenvalue.
\end{SmallList}
\end{coro}

\begin{proof}
As both $H_V$ and $H_{-V}$ have only finitely many eigenvalues
below zero, the solutions $u$ and $v$, as defined in
\eqref{E:uwDefn}, have only finitely many zeros. If we define $x_0
= \max \{ x : u(x) v(x) = 0 \}$, then $u(x)$ and $v(x)$ do not
change sign for $x \ge x_0$. By symmetry, we may suppose that
$u(x_0) = 0$.

Let $V_1(x)=V(x+x_0)$.  As $u$ and $v$ do not change sign for
$x>x_0$, both $H_{V_1}$ and $H_{-V_1}$ have spectrum contained in
$[0,\infty)$.  By the previous theorem, part~(a) follows for $V_1$
and so also for $V$. It also shows that $u$ cannot be square
integrable.

To prove that $w$ is not square integrable, we modify $V_1$ as
follows. Consider $V_2 = V_1 + \lambda \chi_{[0,1]}$.  As
$\sigma(H_{V_1})\subseteq[0,\infty)$, the same is true of
$H_{V_2}$ so long as $\lambda\geq 0$.

Choose $\lambda$ to be the smallest eigenvalue of the following
problem on $[0,1]$:
$$
-\frac{d^2\psi}{dx^2} - V_1 \psi = \lambda \psi, \qquad \psi(0)=0,
\quad \psi'(1)v(x_0+1)-\psi(1)v'(x_0+1)=0.
$$
As $x \mapsto v(x_0 + x)$ does not have a zero in $[0,1]$,
$\lambda$ cannot be negative. We denote the corresponding
eigenfunction by $\psi$, normalized to have $\psi (1) = v(x_0 +
1)$.

For this value of $\lambda$, the function
$$
v_2(x) = \begin{cases} \psi(x) & 0 \le x \le 1 \\ v(x_0 + x) & 1
\le x < \infty \end{cases}
$$
is the Dirichlet solution for the operator $H_{-V_2}$ at energy
zero and it does not change sign. This implies that
$\sigma(H_{-V_2}) \subseteq [0,\infty)$.

We have just seen that for $\lambda$ fixed as above, both
$H_{V_2}$ and $H_{-V_2}$ have spectrum contained in $[0,\infty)$.
By part~(b) of Theorem~\ref{T:contconseq}, $v_2$ cannot be square
integrable, which implies that $v$ cannot be square integrable,
either.
\end{proof}

\section{Absence of Singular Spectrum: The Continuum
Case}\label{S:7}

Our goal in this section is to prove Theorems~\ref{T:Cbiggy}
and~\ref{T:Cbiggyer}; that is, to show that if the negative
spectrum of both $H_V$ and $H_{-V}$ consists of only finitely many
eigenvalues, then both operators have purely absolutely continuous
spectrum on $[0,\infty)$. We have seen above that zero is not an
eigenvalue, so it suffices to consider the open interval
$(0,\infty)$. Absence of singular spectrum in this interval is a
consequence of the following general result whose applicability is
guaranteed by Theorem~\ref{T:contconseq} or
Corollary~\ref{C:Cfmbs}.

\begin{theorem}\label{T:Cspec}
Let $H = -\Delta + V$ be a continuum half-line Schr\"odinger
operator whose potential can be written as $V = W' + Q$ with $W'
\in \ell^\infty (L^2)$, $Q \in L^1$, and
\begin{equation}\label{E:tw2bound}
\int_1^x t [W(t)]^2 \le \tfrac14 \log (x) + C.
\end{equation}
Then the essential support of the absolutely continuous spectrum
of $H$ is $(0,\infty)$ and the spectrum is purely absolutely
continuous on this set.
\end{theorem}

The proof follows the same strategy as the proof of
Theorem~\ref{T:Dspec}, that is, we prove estimates on the
behaviour of generalized eigenfunctions and then use subordinacy
theory.

Proposition~\ref{P:Cpower} will show that the singular part of the
spectral measure, restricted to $(0,\infty)$, does not assign any
weight to sets of Hausdorff dimension zero.

Proposition~\ref{P:Cboundedsol} will show that for all energies in
$(0,\infty)$, with the exception of a set of zero Hausdorff
dimension, all solutions are bounded. This implies that
$(0,\infty)$ is the essential support of the absolutely continuous
spectrum and that any singular spectrum in $(0,\infty)$ must be
supported on a set of zero Hausdorff dimension.

Notice that these two propositions preclude the existence of
singular spectrum in $(0,\infty)$.

As a preliminary observation, we note the following:

\begin{lemma}\label{L:wbounded}
If $W$ is such that $W' \in \ell^\infty (L^2)$ and
\eqref{E:tw2bound} is satisfied, then $W$ is bounded, square
integrable, and obeys the pointwise estimate
\begin{equation}\label{E:wptwbound}
|W(x)| \lesssim \left( \frac{\log x}{x} \right)^{1/4}
\end{equation}
for $x$ large enough. Moreover, $W^4 W' \in L^1$ and $W \in L^p$
for $p \ge 2$.
\end{lemma}

\begin{proof}
By \eqref{E:tw2bound}, the integral of $|W|^2$ over the interval
$[2^l,2^{l+1}]$ is bounded by $C l 2^{-l}$. Summing this over $l$
proves square integrability.

As $W'\in\ell^\infty(L^2)$, there is a constant $C$ such that, for
$|\delta| \le 1$ and $x > 1$,
$$
|W(x + \delta) - W(x)| = \left| \int_0^\delta W'(x + t) \, dt
\right| \le C |\delta|^{1/2}.
$$
Thus,
$$
|W(x + t)| \ge \tfrac12 |W(x)| \quad \text{for} \quad 0 \le |t|
\le T_x = \min \left\{ \tfrac1{4C}|W(x)|^2 , 1 \right\}.
$$
Combining this with \eqref{E:tw2bound} gives
$$
\min \left\{ \tfrac1{8C} |W(x)|^4 , \tfrac12 |W(x)|^2 \right\} \le
 \int_{-T_x}^{T_x} W(x+t)^2 \, dt  \le \frac{\tfrac14 \log (x+1) + c}{x-1},
$$
which implies that $W(x) \to 0$ as $x \to \infty$ and so
\eqref{E:wptwbound}. As this shows that $W \in L^\infty$ and we
know $W \in L^2$, it follows that $W \in L^p$ for $p \ge 2$.

By the Cauchy--Schwarz inequality, $W'\in\ell^\infty(L^2)$,
\eqref{E:tw2bound}, and \eqref{E:wptwbound},
\begin{align*}
\int_n^{n+1} |W(x)^4 W'(x)| \, dx & \lesssim \left(
\int_n^{n+1} |W(x)|^8 \, dx \right)^{1/2}\\
& \le \left( \sup_{n \le x \le n+1} |W(x)|^3 \right) \left(
\int_n^{n+1} |W(x)|^2 \, dx \right)^{1/2}\\
& \lesssim \left( \frac{\log n}{n} \right)^{3/4} \left( \frac{\log
n}{n} \right)^{1/2}.
\end{align*}
As this is summable, we find $W^4 W'\in L^1$.
\end{proof}

As with its discrete analogue, Theorem~\ref{T:Dspec}, the proof of
Theorem~\ref{T:Cspec} rests on the study of solutions of the
corresponding eigenfunction equation for all boundary conditions.

In order to study solutions of
\begin{equation}\label{E:Csol}
- \psi''(x) + V(x) \psi(x) = k^2 \psi(x),
\end{equation}
we use the continuum Pr\"ufer variables, $R(x)$ and $\theta(x)$.
These are defined by
$$
\psi(x) = R(x) \sin (\theta(x)/2), \; \; \psi'(x) = k R(x) \cos
(\theta(x)/2)
$$
and the requirements that $R(x) > 0$ and $\theta$ be continuous
(c.f.~\cite{kls}). They obey the following differential equations:
\begin{gather}
\label{E:PRC} \frac{d\log R(x)}{dx} = \frac{V(x)}{2k} \sin
\theta (x)\\
\label{E:PTC} \frac{d \theta(x)}{dx} = 2k - \frac{V(x)}{k} (1 -
\cos \theta (x)).
\end{gather}

The following lemma isolates the main term in the asymptotics of
the Pr\"ufer amplitude $R(x)$.

\begin{lemma}\label{L:Rbound}
Under the assumptions of Theorem~\ref{T:Cspec},
\begin{equation}\label{E:logrc}
\log \left( \frac{R(x)}{R(0)} \right) = - \int_0^x W(t) \cos
\theta(t) \, dt +  O(1).
\end{equation}
\end{lemma}

\begin{proof}
From \eqref{E:PRC} and $Q \in L^1$, we find
$$
\log \left( \frac{R(x)}{R(0)} \right) = \tfrac{1}{2k} \int_0^x
V(t) \sin \theta (t) \, dt = \tfrac{1}{2k} \int_0^x W'(t) \sin
\theta (t) \, dt + O(1).
$$
Integration by parts, Lemma~\ref{L:wbounded}, and \eqref{E:PTC}
yield
$$
\tfrac{1}{2k} \int_0^x W'(t) \sin \theta (t) \, dt = - \int_0^x
W(t) \cos \theta(t) \left[ 1 - \frac{V(t)}{2k^2} [ 1 - \cos \theta
(t) ] \right] \, dt + O(1),
$$
so that \eqref{E:logrc} will follow once we show
\begin{equation}\label{E:bddrem}
\int_0^x W(t) W'(t) \cos \theta(t) [ 1 - \cos \theta (t) ] \, dt =
O(1).
\end{equation}
Note that $W(t) W'(t) = \frac12 (W(t)^2)'$. Integrating by parts,
and reusing this idea, shows that \eqref{E:bddrem} holds. Along
the way we use the pointwise bound \eqref{E:wptwbound} to control
the boundary terms, $W \in L^p$ for $p \ge 2$ to control integrals
not containing $W'$, and finally $W^4 W' \in L^1$ to control the
integral that contains this term.
\end{proof}

\begin{lemma}\label{L:cosC}
Assume that for all $L>0$,
\begin{equation}\label{E:cosccond}
\sup_{0 \le t \le L} | \phi(x + t) - \phi(x) - 2kt | \to 0 \mbox{ as } x \to \infty.
\end{equation}
Then, for every $\varepsilon > 0$, there is a constant $C$ so that
\begin{equation*}
\int_1^x \frac{\cos^2 \phi(t)}{t} \, dt \le \left[ \tfrac12 +
\varepsilon \right] \log (x) + C.
\end{equation*}
\end{lemma}

\begin{proof}
Let $\varepsilon > 0$ be given. As $2\cos^2 \phi = 1 + \cos 2
\phi$, it suffices to show that
\begin{equation}\label{E:epscoslog}
\int_1^x \frac{\cos 2\phi(t)}{t} \, dt \le  \varepsilon \log (x) +
C.
\end{equation}
For $L$ sufficiently large, say $L > 2/(\varepsilon k)$, we have
$$
\sup_\delta \left| \int_0^L \cos ( 4kt + \delta ) \, dt \right|
\le \tfrac14 \varepsilon L.
$$
For such an $L$ and $x$ large enough, we have
$$
\left| \int_0^L \cos 2 \phi ( x + t )  \, dt \right| \le \tfrac12
\varepsilon L.
$$
Thus, again for $x$ large enough,
$$
\left| \int_0^L \frac{\cos 2\phi(x + t)}{x + t} \, dt \right| \le
\frac{1}{x} \left| \int_0^L \cos 2 \phi ( x + t )  \, dt \right| +
\int_0^L \frac{t \, dt}{x(x + t)} \le \frac{\varepsilon L}{x}.
$$
From this, \eqref{E:epscoslog} follows by breaking the integral
over $[0,x]$ into $L$-sized blocks.
\end{proof}

\begin{prop}\label{P:Cpower}
Suppose $V = W' + Q$ with $Q \in L^1$ and $W' \in \ell^\infty
(L^2)$ obeying \eqref{E:tw2bound}. Then, for $k > 0$, all
solutions $\psi$ of \eqref{E:Csol} that are not identically zero
obey
$$
x^{-\eta} \lesssim |\psi(x)|^2 + |\psi'(x)|^2 \lesssim x^\eta
$$
for any $\eta > 1/\sqrt{2}$ and $x \ge 1$. Consequently, the
spectral measure gives zero weight to any subset of $(0,\infty)$
of Hausdorff dimension less than $1 - 2^{-1/2}$.
\end{prop}

\begin{proof}
Fix $\eta > 1/\sqrt{2}$. By Lemma~\ref{L:Rbound}, it suffices to
show
$$
\left| \int_1^x W(t) \cos \theta(t) \, dt \right| \le
\frac{\eta}{2} \log (x) + O(1).
$$
By the Cauchy--Schwarz inequality,
$$
\left| \int_1^x W(t) \cos \theta(t) \, dt \right|^2 \le \int_1^x t
W(t)^2 \, dt \cdot \int_1^x \frac{\cos^2 \theta(t)}{t} \, dt.
$$
Therefore, once we show that the function $\theta$ satisfies the
condition \eqref{E:cosccond}, Lemma~\ref{L:cosC} and
\eqref{E:tw2bound} allow us to conclude the proof. To this end, we
note that
$$
\theta(x+t) - \theta(x) - 2kt = - \tfrac{1}{k} \int_0^t V(x+s) [1
- \cos \theta(x+s)] \, ds
$$
and hence
\begin{align*}
\sup_{0 \le t \le L} \left| \theta(x+t) - \theta(x) - 2kt \right|
\le &  \sup_{0 \le t \le L} \left|\tfrac{1}{k} \int_0^t W'(x+s) [
1 - \cos \theta(x+s) ]
\, ds \right|\\
& \qquad + \tfrac{2}{k} \int_0^L |Q(x+s)| \, ds.
\end{align*}
As $Q \in L^1$, the second term goes to zero as $x \to \infty$. To
show that
$$
\sup_{0 \le t \le L} \left| \int_0^t W'(x+s) [ 1 - \cos
\theta(x+s) ] \, ds \right| \to 0 \mbox{ as } x \to \infty,
$$
we integrate by parts four times, as in the proof of
Lemma~\ref{L:Rbound}, and then apply Lemma~\ref{L:wbounded}.

The statement about the spectral measure follows from the
Jitomirskaya-Last version of subordinacy theory \cite{jl}.
\end{proof}

Our next goal is to show that the set of energies at which not all
solutions of \eqref{E:Csol} are bounded is of zero Hausdorff
dimension. First we prove a continuum analogue of
Lemma~\ref{klstool}; see \cite[Theorem~3.2]{kls} for a related
result.

\begin{lemma}\label{L:Cklstool}
Suppose $V = W' + Q$ with $W$ and $Q$ as above.  Fix $k \in
(0,\infty)$. If
\begin{equation}\label{whatever2}
\widehat{W}(k;x) \equiv \lim_{M \to \infty} \int_{x}^M W(t)
e^{2ikt} \, dt
\end{equation}
exists and obeys
\begin{equation}\label{E:Ckls}
\widehat{W} W \in L^1,
\end{equation}
then all solutions of \eqref{E:Csol} are bounded.
\end{lemma}

\begin{proof}
Let $k$ be such that  $\widehat{W}(k;x)$ exists and \eqref{E:Ckls}
holds. By Lemma~\ref{L:Rbound}, it suffices to show that $\int_0^x
W(t) e^{i \theta(t)} \, dt$ is bounded. Notice that the existence
of the limit in \eqref{whatever2} implies that
\begin{equation}\label{E:whatzero}
\widehat{W}(k;x) \to 0 \mbox{ as } x \to \infty
\end{equation}
and, by \eqref{E:wptwbound},
\begin{equation}\label{E:wzero}
W(x) \to 0 \mbox{ as } x \to \infty.
\end{equation}
Now we proceed as follows:
\begin{align*}
\int_0^x W(t) e^{i \theta(t)} \, dt &= \int_0^x
\frac{\partial}{\partial t} \widehat{W}(k;t) e^{i \theta(t) -
2ikt} \, dt\\
&= \tfrac{i}{k} \int_0^x \widehat{W}(k;t) V(t) [ 1 - \cos
\theta(t)
] e^{i \theta(t) - 2ikt} \, dt + O(1)\\
&= \tfrac{i}{k} \int_0^x \widehat{W}(k;t) W'(t) e^{- 2ikt}
P(\theta(t)) \, dt + O(1),
\end{align*}
where we used \eqref{E:whatzero} and \eqref{E:wzero} in the second
step and $Q \in L^1$ in the last step. Here, $P(\cdot)$ denotes a
trigonometric polynomial. Integrating by parts four times shows
that this integral is bounded because $W^4 W' \in L^1$. To make
this more explicit, one may use the following observation four
times (with $l = 0$, $1$, $2$, and then $3$): Given $l \ge 0$ and
a trigonometric polynomial $P_1(t,\theta)$, there is a
trigonometric polynomial $P_2(t,\theta)$ such that
$$
\int_0^x \widehat{W}(k;t) W^l (t) W'(t) P_1(t,\theta(t)) \, dt  =
\int_0^x \widehat{W}(k;t) W^{l+1} (t) W'(t) P_2(t,\theta(t)) \, dt
+ O(1).
$$
This is proved by integration by parts:
\begin{align*}
\int_0^x & \widehat{W}(k;t) W^l (t) W'(t) P_1(t,\theta(t)) \, dt =
\int_0^x \widehat{W}(k;t) \frac{(W^{l+1} (t))'}{l+1}
P_1(t,\theta(t)) \, dt \\
&= - \tfrac{1}{l+1} \int_0^x \widehat{W}(k;t) W^{l+1} (t) \left[
\tfrac{\partial}{\partial t} P_1(t,\theta(t)) +
\tfrac{\partial}{\partial \theta}
P_1(t,\theta(t)) \theta'(t) \right] \, dt\\
& \hspace{6mm} - \tfrac{1}{l+1} \int_0^x W^{l+2} (t) e^{2ikt}
P_1(t,\theta(t)) \, dt + O(1)\\
&= \int_0^x \widehat{W}(k;t) W^{l+1} (t) W'(t) P_2(t,\theta(t)) \,
dt + O(1).
\end{align*}
Here we used \eqref{E:whatzero}, \eqref{E:wzero}, $W \in L^2$ (see
Lemma~\ref{L:wbounded}), and the assumption \eqref{E:Ckls}, though
only in the case $l=0$.
\end{proof}

To use this lemma to show that the set of energies at which not
all solutions of \eqref{E:Csol} are bounded is of zero Hausdorff
dimension, we need to control $\widehat{W}$. For this, we use the
following analogue of Lemma~\ref{L:S-Z} whose proof is a
straightforward adaptation of the arguments in \cite[\S
XIII.11]{Zygmund} or \cite[\S V.5]{Carleson}. The two ingredients
will be combined in Proposition~\ref{P:Cboundedsol} below.

\begin{lemma}\label{L:CS-Z}
For each $\varepsilon \in (0,1)$, every measurable function $m :
(0,\infty) \to \R$, and every measure $\nu$, we have
$$
\biggl\{ \int\ \biggl| \int_0^{m(k)} g(t) e^{2ikt} \, dt \biggr|
\, d\nu(k) \biggr\}^2 \lesssim \mathcal{E}_\varepsilon (\nu) \int
(1+t^2)^{\frac{1-\varepsilon}{2}} |g(t)|^2 \, dt,
$$
where $\mathcal{E}_\varepsilon (\nu) = \int \int ( 1 +
|x-y|^{-\varepsilon}) \, d\nu(x) \, d\nu(y)$ denotes the
$\varepsilon$-energy of $d\nu$.
\end{lemma}

\begin{prop}\label{P:Cboundedsol}
Suppose $V = W' + Q$ with $Q \in L^1$ and $W' \in \ell^\infty
(L^2)$ obeying \eqref{E:tw2bound}. There is a set $S \subseteq
(0,\infty)$ of zero Hausdorff dimension so that for all $E \in
(0,\infty) \setminus S$, all solutions $\psi$ of \eqref{E:Csol}
are bounded. Consequently, the singular part of the spectral
measure on $(0,\infty)$ is supported by a set of zero Hausdorff
dimension.
\end{prop}

\begin{proof}
The proof is completely analogous to the proof of
Proposition~\ref{P:boundedsol}, so we just sketch the argument.
Let $m(k)$ be a measurable function and for every $l \ge 0$, let
$m_l(k) = \max \{ 2^l, m(k) \}$ and $\Omega_l = \{ k : m(k) \le
2^{l+1} \}$. Then, it follows from Lemma~\ref{L:CS-Z} that for
every $\varepsilon \in (0,1)$,
$$
\int_{\Omega_l} \biggl| \int_{m_l(k)}^{2^{l+1}} 2^{\varepsilon
l/4} W(t) e^{2ikt} \, dt \biggr| \, d\nu(k) \lesssim
\sqrt{\mathcal{E}_\varepsilon (\nu)}\, 2^{-\varepsilon l/4}
\sqrt{l}.
$$
This shows that the set of $k$ for which $x^{\varepsilon/4}
\widehat{W}(k;x)$ is unbounded must be of zero
$\varepsilon$-capacity, and hence of Hausdorff dimension no more
than $\varepsilon$. Since $x^{-\varepsilon/4} W(x) \in L^1$, an
application of Lemma~\ref{L:Cklstool} completes the proof of the
proposition.

The last statement follows from the well-known fact that the
spectral measure is purely absolutely continuous on the set of
energies where all solutions are bounded \cite{gp,s,stolz}.
\end{proof}

\end{document}